\documentclass[%
reprint,
 amsmath,amssymb,
 aps,
]{revtex4-2}

\usepackage{subfigure}
\usepackage{float}
\usepackage{slashed}
\usepackage{xcolor}
\usepackage{amsmath}
\usepackage{subcaption}
\usepackage[english]{babel}
\usepackage{tikz}
\usepackage{adjustbox}
\usepackage{rotating}
\usepackage{rotfloat}
\usepackage{verbatim}
\usepackage[utf8]{inputenc}
\usepackage{graphicx}
\usepackage{dcolumn}
\usepackage{bm}
\usepackage[super]{nth}
\usepackage{lipsum}

\newcommand\Tdiag[4]{%
    \multicolumn{1}{|p{#2}|}{\hskip-\tabcolsep
    \begin{tikzpicture}[%
                baseline={(0,-.25\baselineskip)},
                every node/.style={outer sep=0pt,inner sep=#1}]
    \node[minimum width={#2+1\tabcolsep-\pgflinewidth},
        minimum height=2\baselineskip-\pgflinewidth+\extrarowheight,
        use as bounding box] (box) {};
    \draw[line cap=round] (box.north west) -- (box.south east);
    \node[anchor=south west,text width=.75*#2,align=left] at (box.south west) {#3};
    \node[anchor=north east,text width=.75*#2,align=right] at (box.north east) {#4};
\end{tikzpicture}\hskip-\tabcolsep}}

\usepackage{hyperref}
\usepackage{xcolor}
\hypersetup{
  colorlinks   = true, 
  urlcolor     = red, 
  linkcolor    = blue, 
  citecolor   = blue 
}

\usepackage{threeparttable}
\begin{document}

\preprint{APS/123-QED}

\title{Probing torsion field with Einstein-Cartan theory at the HL-LHC: an angular distribution case study}

\author{S. Elgammal}
 \altaffiliation[sherif.elgammal@bue.edu.eg]{}
\affiliation{%
Centre for Theoretical Physics, The British University in Egypt, P.O. Box 43, El Sherouk City, Cairo 11837, Egypt.
}%


\date{\today}

\begin{abstract}
{This analysis utilizes simulated data privately generated based on the High Luminosity Large Hadron Collider (HL-LHC) configuration to investigate the angular distribution of high-mass dimuon pairs produced during the foreseen proton-proton collisions at a center-of-mass energy of 14 TeV. 
The study focuses on the cos$\theta_{CS}$ variable, which is defined in the Collins-Soper frame. In the Standard Model, the production of high-mass dimuon pairs is primarily governed by the Drell-Yan process, which demonstrates a significant forward-backward asymmetry.
However, scenarios beyond the Standard Model suggest different shapes for the angular distribution (cos$\theta_{CS}$). By observing excess events not predicted by the Standard Model, the angular distribution can help differentiate among these alternative models. 
Furthermore, we used a simplified Einstein-Cartan model to analyze the simulated data. This analysis established upper limits at the 95\% confidence level regarding the masses of various particles within the model, including a spin-2 dark neutral gauge boson and the torsion field.}

\vspace{0.75cm}
\end{abstract}

\maketitle






\section{Introduction}
\label{sec:intro}
One of the methods for detecting physics beyond the Standard Model (SM) \cite{SMcource} at the Large Hadron Collider (LHC) involves observing changes in the dilepton mass spectrum at high mass. Such changes might appear as a new peak, as anticipated by heavy neutral gauge boson models, like Z$^{\prime}$ \cite{heaveyZ}, or by Randall-Sundrum models \cite{extradim}. Alternatively, the spectrum could display a broad distortion, which might indicate the presence of Contact Interactions \cite{leptoquark, ContactInteraction} or models such as ADD \cite{ADD}.

The CMS collaboration has focused extensively on these signatures, particularly concerning Z$^{\prime}$ and Contact Interaction models \cite{ZprimeandCI}. Another model, known as Mono-Z$^{\prime}$ \cite{monoZprime, monoZprime2}, predicts the production of dark matter in conjunction with Z$^{\prime}$.

Previous studies performed by the CMS \cite{AfbCMS} and ATLAS \cite{AfbATLAS} collaborations utilized the angular distributions of Drell-Yan charged lepton pairs around the Z boson mass peak to measure the forward-backward asymmetry \( (A_{FB}) \). Both studies analyzed the complete LHC Run 1 data, which included an integrated luminosity of 19.7 fb\(^{-1}\) for CMS and 20.3 fb\(^{-1}\) for ATLAS, based on proton-proton collisions at a center of mass energy of 8 TeV \((\sqrt{s})\). The measurements of \( A_{FB} \) were found to be consistent with predictions from the Standard Model.

Moreover, the forward-backward asymmetry of high-mass dileptons (with \( M_{ll} > 170 \) GeV) has been measured using the CMS detector at \( \sqrt{s} = 13 \) TeV, with an integrated luminosity of 138 fb\(^{-1}\). This analysis concluded that no statistically significant deviations from the Standard Model predictions were observed \cite{AfbCMS13tev}.

In the current analysis, we studied the angular distributions of the dimuon channel in mass bins above the Z-boson mass peak, utilizing Monte Carlo simulated data from the CMS experiment for the HL-LHC \cite{R14}, which has an integrated luminosity of 3000 $\text{fb}^{-1}$ at a center-of-mass energy of $\sqrt{s} = 14$ TeV \cite{BSM-HL-LHC}. 
To interpret the simulated data, we employed a simplified model \cite{R1} based on Einstein-Cartan (EC) theory \cite{EC1,EC2,EC3,EC4,EC5,EC6}. 
This model is for the dark neutral gauge boson (A$^{\prime}$) production alongside dark matter (DM) at the HL-LHC. Many cosmological observations, including recent measurements from \cite{planck2015, planck2018, bullet_cluster}, support the existence of dark matter. These observations suggest that dark matter is non-decaying, and weakly interacting massive particles \cite{R6,Maverick_dark_matter_at_colliders}.

Several searches for dark matter (DM) have been conducted using data collected from the CMS and ATLAS experiments during RUN 2 \cite{CMS-DM, ATLAS-DM}. These searches focus on the production of a visible object, referred to as (X), which recoils against the significant missing transverse energy resulting from dark matter particles. This creates a signature of \((\text{X} + E^{\text{miss}}_{T})\) in the detector \cite{R38}. The visible particle "X" could be a Standard Model (SM) particle, such as W or Z bosons, jets \cite{R35, atlasmonoZ, mono-zCMS, mono-zATLAS}, a photon \cite{photon, photonATLAS}, or even the SM Higgs boson \cite{R36, monoHiggsAtlas1, monoHiggsAtlas2}. 

Additionally, we provide a statistical interpretation of the results. The simulated data for the signal and standard-model backgrounds were produced privately.

The analysis is structured as follows:
A theoretical model based on Einstein-Cartan theory is discussed in section \ref{section:model}. 
Section \ref{section:CS} introduces the Collins-Soper frame and the variable cos$\theta_{\text{CS}}$ as it pertains to Drell-Yan events, including the quark direction ambiguity. 
A brief overview of the HL-LHC project is provided in section \ref{section:CMS}. 
Section \ref{section:Backgrounds} covers the Monte Carlo production of signal samples used in the analysis, followed by an exploration of significant SM background processes.
The analysis strategy and event selection criteria are detailed in section \ref{section:AnSelection}. In section \ref{section:Results}, we present results based on angular distribution (cos$\theta_{\text{CS}}$) and statistical interpretation. Finally, a summary is offered in section \ref{section:Summary}.
\section{The theoretical signal model}
\label{section:model}
\begin{figure} [h!]
\centering
\resizebox*{6.0cm}{!}{\includegraphics{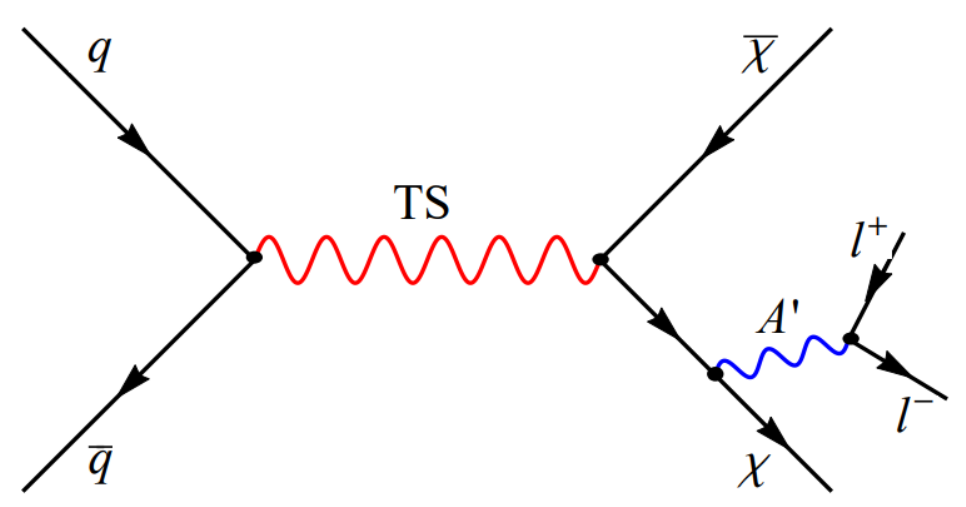}}
\caption{The Feynman diagram for quark anti-quark interactions mediates the torsion field (TS), which transforms into two dark matter particles (\(\chi\bar{\chi}\)). One of these particles radiates a dark gauge boson ($A'$) via bremsstrahlung \cite{R1}.}
\label{fig1}
\end{figure}
The simplified model analyzed is based on Einstein-Cartan theory, which discusses dark matter production from proton-proton collisions at the HL-LHC, alongside a new neutral dark gauge boson A$^{\prime}$ \cite{R1}. 

This dark gauge boson (A$^{\prime}_{\mu}$) can result from the pair annihilation of quarks $q\bar{q}$ mediated by the heavy torsion field ($TS$), leading to the generation of two dark matter particles ($\chi$). The dark matter is sufficiently heavy to decay into an A$^{\prime}_{\mu}$ and another dark matter particle ($\chi$), as illustrated in figure \ref{fig1}.

The interaction terms, in the effective Lagrangian, between the torsion field 
and Dirac fermion $(\psi)$, is given by \cite{R1} 
\begin{equation}
\bar{\psi}i\gamma^{\mu}(\partial_{\mu} + i\texttt{g}_{\eta}\gamma^{5}S_{\mu} + ...)\psi. \nonumber
\end{equation}

According to the previous equation, the torsion field interacts with both the SM and hidden fermions through its axial-vector component, characterized by a universal coupling constant of \( \texttt{g}_{\eta} \). 
In dynamical torsion theories, such as the one discussed in \cite{R1}, the classical coupling constant \( \texttt{g}_{\eta} \) holds a fixed value of \( \frac{1}{8} \) (0.125). This specific value emerges from the underlying structure of the theory and how torsion interacts with fermions.

The effective Lagrangian term for the torsion field coupling to dark matter and the interaction between the dark gauge boson (A$^{\prime}_{\mu}$) and dark matter is given by \cite{R1}.
\begin{equation}
\bar{\chi}(i\gamma^{\mu}D_{\mu} - M_{\chi})\chi. \nonumber
\end{equation}

The expression for the covariant derivative is \(D_{\mu} = \partial_{\mu} + i\texttt{g}_{\eta}\gamma^{5}S_{\mu} + i\texttt{g}_{D} A^{\prime}_{\mu}\). The dark matter mass is denoted by \(M_{\chi}\), and the dark gauge boson coupling to dark matter is \(\texttt{g}_{D} = 1.0\) as suggested from the LHC Dark Matter Working Group \cite{DMrecommendations}.

\begin{table*}
\fontsize{10.pt}{12pt}
\selectfont
\caption{The parameters of the simplified model drawing on the principles of Einstein-Cartan theory, all while considering a specific mass assumption \cite{R1}.}
\begin {tabular} {lll}
\hline
\hline
Parameters & \hspace{1pt} Description & \hspace{1pt} optimized value \\
\hline
$M_{TS}$  & Mass of torsion field $(\text{TS})$ & \hspace{10pt} $[1250,...,7000]$ GeV\\
$M_{A'}$  & Mass of dark gauge boson $(A')$ & \hspace{10pt} $[200,...,600]$ GeV\\
$M_{\chi}$ & Mass of dark matter $(\chi)$ & \hspace{10pt} $500$ GeV \cite{indirect2, Nature-indirect}\\
$\texttt{g}_{D}$ & Coupling of dark gauge boson to dark matter & \hspace{10pt} 1.0 \cite{DMrecommendations} \\
$\texttt{g}_{\eta}$  & Coupling of torsion field to Dirac fermions & \hspace{10pt} 0.125 \cite{R1}\\
\hline
Mass assumption & \hspace{4pt} $M_{A^{\prime}} < 2M_{\chi}$ \cite{R1} \\
\hline
\hline
\end {tabular}
\vspace{3pt}
\label{table:par}
\end{table*}

\begin{sidewaystable} []
\caption{The simplified model (based on Einstein-Cartan theory) cross-section measurements times branching ratios (in pb) calculated for different sets of the masses $M_{A^{\prime}}$ (in GeV), and $M_{TS}$ (in GeV), for the mass assumption given in table \ref{table:par}, with dark matter mass ($M_{\chi} = 500$ GeV), the following couplings constants $\texttt{g}_{\eta} = 0.125,~\texttt{g}_{D} = 1.0$ and at $\sqrt{s} = 14$ TeV.}
\fontsize{9.pt}{12pt}
\selectfont
\begin{tabular}{|c|c|c|c|c|c|c|c|c|c|c|c|}
\hline
\Tdiag{.4em}{1.8cm}{$M_{A'}$}{$M_{TS}$}  & 1250 &1500 &1750 & 1800 & 1970 & 2000 & 3000 & 4000 & 5000 & 6000 & 7000 \\
\hline
200 &  7.1$\times10^{-5}$ & 
       10.0$\times10^{-4}$ & 
       26.7$\times10^{-4}$ & 
       28.2$\times10^{-4}$ & 
       30.6$\times10^{-4}$ &
       30.8$\times10^{-4}$ &
       15.8$\times10^{-4}$ &
       4.7$\times10^{-4}$ &
       1.2$\times10^{-4}$ &
       2.6$\times10^{-5}$ &
       5.1$\times10^{-6}$ \\

\hline
300 &  1.3$\times10^{-5}$ & 
       6.6$\times10^{-4}$ & 
       18.0$\times10^{-4}$ & 
       19.7$\times10^{-4}$ & 
       23.7$\times10^{-4}$ &
       24.0$\times10^{-4}$ &
       15.1$\times10^{-4}$ &
       4.8$\times10^{-4}$ &
       1.3$\times10^{-4}$ &
       2.8$\times10^{-5}$ &
       5.6$\times10^{-6}$ \\
\hline
400 &  9.8$\times10^{-7}$ & 
       1.7$\times10^{-4}$ & 
       10.8$\times10^{-4}$ & 
       12.4$\times10^{-4}$ & 
       16.9$\times10^{-4}$ &
       17.4$\times10^{-4}$ &
       13.7$\times10^{-4}$ &
       4.7$\times10^{-4}$ &
       1.3$\times10^{-4}$ &
       2.9$\times10^{-5}$ &
       6.0$\times10^{-6}$ \\
\hline
500 &  7.5$\times10^{-7}$ & 
       9.7$\times10^{-7}$ & 
       2.7$\times10^{-4}$ & 
       3.5$\times10^{-4}$ & 
       7.7$\times10^{-4}$ &
       7.9$\times10^{-4}$ &
       12.3$\times10^{-4}$ &
       4.6$\times10^{-4}$ &
       1.3$\times10^{-4}$ &
       3.0$\times10^{-5}$ &
       6.3$\times10^{-6}$ \\
\hline
600 &  6.1$\times10^{-7}$ & 
       7.7$\times10^{-7}$ & 
       1.8$\times10^{-4}$ & 
       2.9$\times10^{-4}$ & 
       6.4$\times10^{-4}$ &
       7.0$\times10^{-4}$ &
       10.1$\times10^{-4}$ &
       4.1$\times10^{-4}$ &
       1.2$\times10^{-4}$ &
       2.8$\times10^{-5}$ &
       6.0$\times10^{-6}$ \\
\hline
\end {tabular}
\label{table:tabchi}
\end{sidewaystable}

The neutral dark gauge boson (A$^{\prime}$) decays into SM fermion pairs, specifically focusing on the muonic decay A$^{\prime} \rightarrow \mu^{+}\mu^{-}$, which has the highest branching ratio under certain mass conditions outlined in table \ref{table:par}. The model includes several free parameters: the masses of the torsion field ($M_{TS}$), dark gauge boson ($M_{A^{\prime}}$), and dark matter ($M_{\chi}$). 

Previous indirect searches for dark matter have shown a promising chance of detecting dark matter with masses ranging from 100 to 10,000 GeV \cite{indirect2, Nature-indirect}. For our analysis, we have chosen the dark matter mass to be \(M_{\chi} = 500\) GeV. In this context, we have fixed the coupling constants at \(\texttt{g}_{\eta} = 0.125\) and \(\texttt{g}_{D} = 1.0\). Meanwhile, the parameters \(M_{TS}\) and \(M_{A^{\prime}}\) are treated as free variables, and their specific values are detailed in Table \ref{table:par}.

The signature of this process features a pair of opposite-sign muons from the decay of A$^{\prime}$, along with significant missing transverse energy due to stable dark matter $\chi$. We focus on the topology $\mu^{+}\mu^{-} + E^{miss}_{T}$, which offers a clean signal against SM backgrounds, especially since the CMS detector is optimized for this channel. 
\section{The Collins-Soper frame}
\label{section:CS}

In hadron colliders, two partons can collide to produce a lepton pair ($l^+l^-$) through the Drell-Yan process ($q\bar{q} \rightarrow \gamma^*/Z \rightarrow l^+l^-$), which is the only tree-level process in the Standard Model for this outcome. 
In the current analysis, we analyze the angular distribution of these pairs in the Collins-Soper (CS) frame \cite{CSpaper}, which reduces distortions from the transverse momenta of the incoming partons. However, accurately measuring the angle $\theta$ between the negative lepton and one of the partons is challenging if either parton has non-zero transverse momentum, and the direction of the quark is not easily determined in $q\bar{q}$ annihilation.

We adopt the Collins-Soper frame to explore the angular distribution of lepton pairs effectively. In this frame, the angle $\theta_{CS}$ is defined as the angle between the momentum of the negative lepton and the z-axis. By using this frame, we aim to reduce the uncertainties that arise from the unknown transverse momentum of the incoming quarks.

To determine the orientation of the Collins-Soper frame, we rely on the sign of the longitudinal boost of the dilepton system. We can compute the angle $\text{cos} \theta_{CS}$ from quantities that we measure in the lab frame, as explained in \cite{AfbCMS}.
\begin{equation}
     \text{cos}\theta_{CS} = \frac{|Q_z|}{Q_z} \frac{2(P_1^+ P_2^- - P_1^- P_2^+)}{\sqrt{Q^2(Q^2 + Q_T^2)}}.     
    \label{costheta:equ}
\end{equation}
The symbols $Q$, $Q_T$, and $Q_z$ stand for the four-momentum, the transverse momentum, and the longitudinal momentum of the dimuon system, respectively. Similarly, $P_1$ ($P_2$) represents the four-momentum of $\mu^-$ ($\mu^+$), and $E_i$ denotes the energy of the muon. In addition, $P^\pm_i$ is defined as $(E_i \pm P_{z,i})/\sqrt{2}$.
\section{The HL-LHC Project}
\label{section:CMS}

The LHC has undergone several upgrades, including Runs I, II, and III, proving its value. The upcoming HL-LHC upgrade will enhance the center of mass energy ($\sqrt{s} = 14$ TeV) and increase proton collisions for better data. This will involve installing new equipment over 1.2 km of the 27 km LHC \cite{R14}.

The CMS detector at CERN's LHC searches for new physics using a complex structure and various angular measurements: the polar angle ($\theta$), azimuthal angle ($\phi$), and pseudo-rapidity ($\eta$). The coordinate system has the z-axis along the beam axis, the x-axis toward the LHC center, and the y-axis upwards. The $\phi$ measurement occurs in the x-y transverse plane, while $\theta$ is along the x-axis. The direction of the collision products is described by $\eta$, defined as $\eta = - \text{ln}[\text{tan}(\theta/2)]$ \cite{R27,R28}.
\section{Simulated samples and background estimation}
\label{section:Backgrounds}
\subsection{Monte Carlo simulation of the model signals}
We generated the events of the signal model using $\text{MadGraph5}\_\text{aMC@NLO v3.5.0}$ \cite{MG5}. The cross-section was calculated at NLO, and Pythia 8 \cite{R34} was employed for the hadronization process, and DELPHES \cite{delphes} for a fast detector simulation of HL-LHC. We analyzed the production cross-section for various mass combinations of the dark gauge boson ($M_{A^{\prime}}$) and torsion field ($M_{TS}$), assuming $\texttt{g}_{\eta} = 0.125$, $\texttt{g}_{D} = 1.0$, and a dark matter mass of $M_{\chi} = 500$ GeV. The Monte Carlo signal samples used in this analysis and their corresponding cross-sections are listed in Table \ref{table:tabchi}.
\subsection{Monte Carlo simulation of the SM backgrounds}

The SM background processes yielding muon pairs in the signal region are 
Drell-Yan ($\text{DY}\rightarrow \mu^+\mu^-$), the production of top quark pairs ($\text{t}\bar{\text{t}} \rightarrow \mu^+\mu^- + 2b + 2\nu$), $tW \rightarrow \mu^+\mu^- + 2\nu +b$, and production of diboson 
($W^{+}W^{-} \rightarrow \mu^+\mu^- + 2\nu$,  
$ZZ \rightarrow \mu^+\mu^- + 2\nu$ and 
$W^{\pm}Z \rightarrow \mu^\pm \mu^+\mu^- + \nu$).

Another source of background is the jet background, which occurs when jets are misidentified as muons, especially in W+jets and QCD multijet processes. This background is usually estimated using a data-driven method, as detailed in reference \cite{zprime}. However, it is not relevant to our study, since the yield from these processes has been measured to be among the lowest of the standard model backgrounds according to reference \cite{zprime}.

In addition, further cuts are implemented, detailed in section \ref{section:Results}, which rely on variables such as ($\Delta\phi$(lepton, MET), MET, etc.). These cuts are often used to strongly suppress the QCD events in which MET aligns with a mismeasured jet, as discussed in ref. \cite{QCD}.

The SM processes were generated using \text{MadGraph5\_aMC@NLO} with Pythia 8 for parton showering and DELPHES for fast detector simulation of the HL-LHC. They were derived from 14 TeV proton-proton collisions, with muon $p_{T} > 10$ GeV and $|\eta| < 3$.

All Monte Carlo samples used in this analysis and their cross sections were computed in next-to-leading order. Signal samples and SM background contributions were estimated from the simulations, normalized to their corresponding cross-section, and an integrated luminosity of 3000 fb$^{-1}$.

\section{Event selection}
\label{section:AnSelection}
The event selection for the analysis aims to reconstruct a final state with two high transverse momentum $(p_{T})$ muons and missing transverse energy, indicative of a dark matter candidate. 

The pre-selection criteria for each muon include:  
$p^{\mu}_{T}$ $> 30$ GeV,  
$|\eta^{\mu}|$ $<$ 2.5,  
and isolation variable $\Sigma_{i} p^{i}_{T}/p^{\mu}_{T} < 0.1$.  

The isolation cut in DELPHES requires that the scalar $p_{T}$ sum of all muon tracks within a cone of $\Delta R = 0.5$ around the muon does not exceed 10\% of the muon's $p_{T}$, adjusted for pileup effects.

Events are selected with two oppositely charged muons, and the dimuon invariant mass must exceed 60 GeV to search for high-mass resonances. These cuts are summarized in Table \ref{table:selection2}.
\begin{figure}[h]
\centering
\resizebox*{9.5cm}{!}{\includegraphics{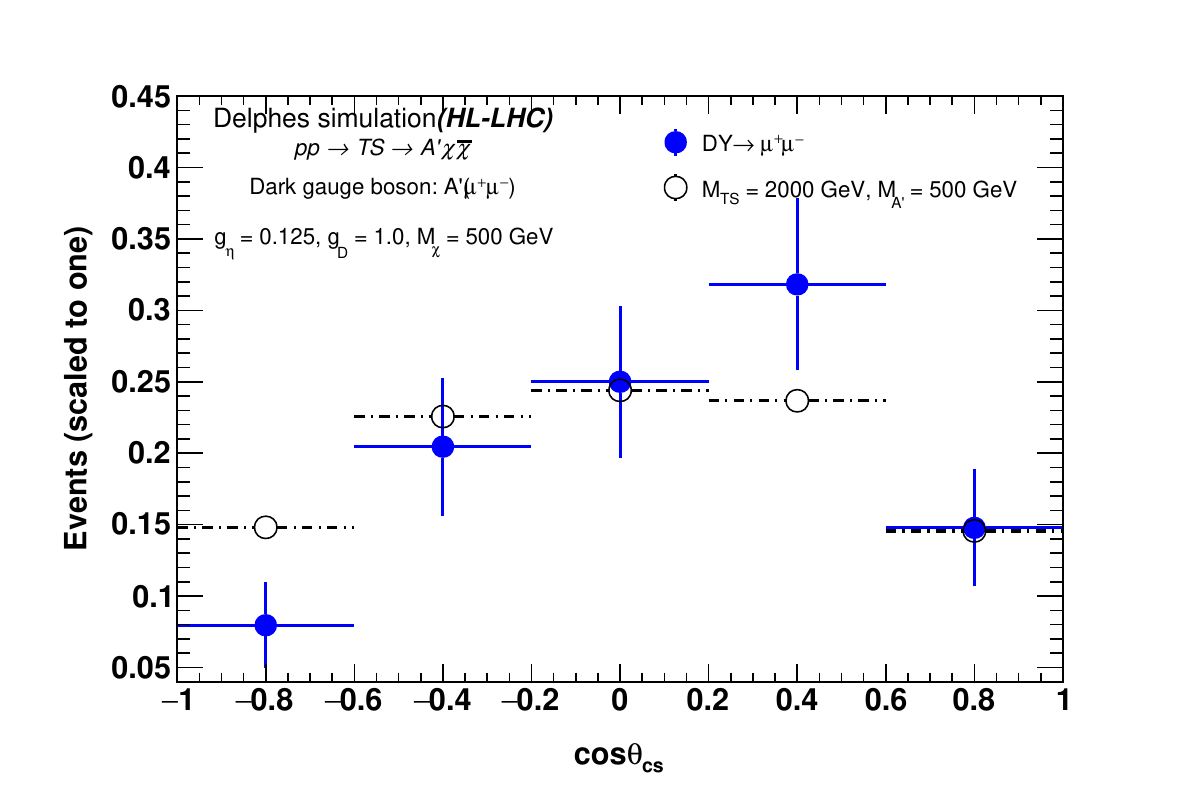}}
\caption{Normalized distributions of cos$\theta_{CS}$ for one resonant model based on Einstein-Cartan theory, generated with a mass of $A^{\prime}$ = 500 GeV, and Drell-Yan events. 
All events must pass the pre-selection listed in table \ref{table:selection2} and have a reconstructed invariant mass in the 460 - 540 GeV range.
All histograms are normalized to unity to highlight qualitative features.} 
\label{fig2}
\end{figure}

\begin{figure}[h]
\centering
\resizebox*{9.5cm}{!}{\includegraphics{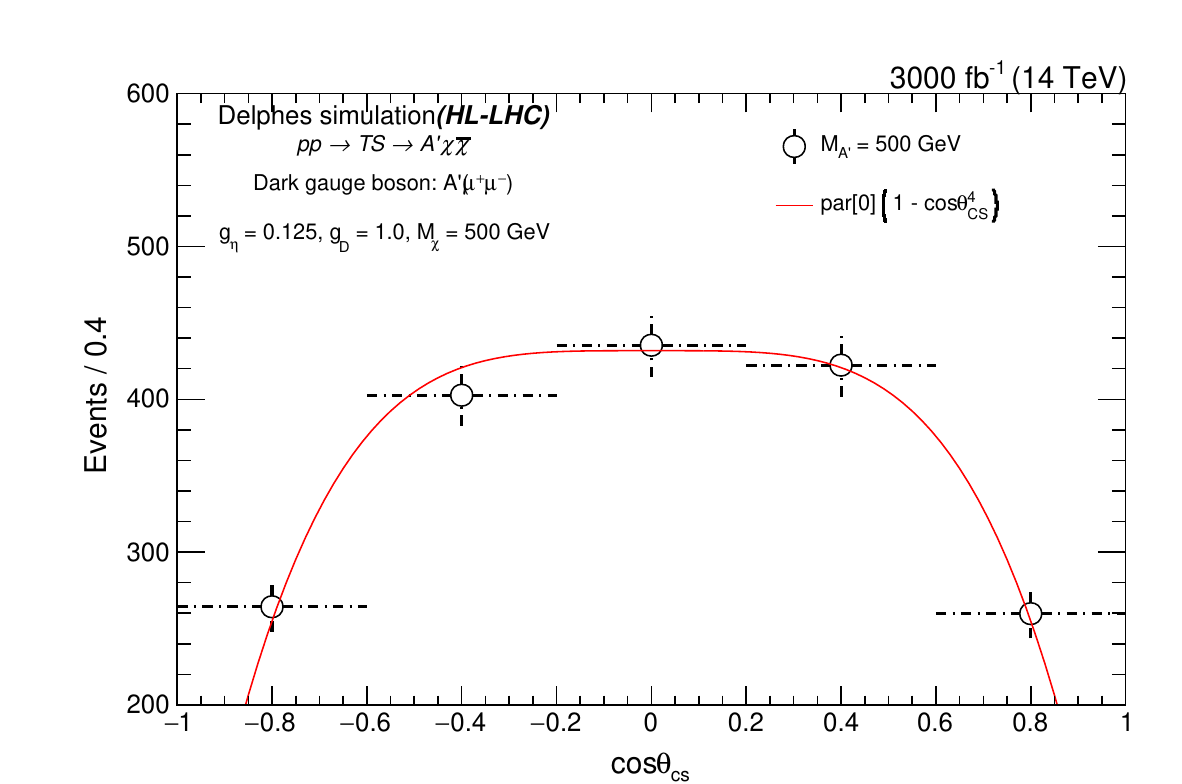}}
\caption{The distributions of cos$\theta_{CS}$ for one resonant model based on Einstein-Cartan theory, generated with a mass of $A^{\prime}$ = 500 GeV at $\sqrt{s} $ = 14 TeV. 
All events must pass the pre-selection listed in table \ref{table:selection2} and have a reconstructed invariant mass in the 460 - 540 GeV range.
The MC data are fitted by the function $\text{par[0]} \big(1 - \text{cos}\theta^{4}_{CS} \big)$
presented by red curve.} 
\label{figure-spin}
\end{figure}

\begin{table}
    \caption{Summary of cut-based event selections used in the analysis.}
    \begin{tabular}{|c|c|}
\hline
Pre-selection & Final selection  \\
\hline
    \hline
  $p^{\mu}_{T} >$ 30 GeV &  $p^{\mu}_{T} >$ 30 GeV   \\
  $|\eta^{\mu}| <$ 2.5   & $|\eta^{\mu}| <$ 2.5 \\
  $\Sigma_{i} p^{i}_{T}/p^{\mu}_{T} < 0.1$ &  $\Sigma_{i} p^{i}_{T}/p^{\mu}_{T} < 0.1$\\
  $M_{\mu^{+}\mu^{-}} > 60$ GeV& $M_{A^{\prime}} - 40 < M_{\mu^{+}\mu^{-}} < M_{A^{\prime}} + 40$\\
  &$\Delta\phi_{\mu^{+}\mu^{-},\vec{E}_{T}^{\text{miss}}} >$ 2.5 \\
  &$|E_{T}^{\mu^{+}\mu^{-}}- E_{T}^{\text{miss}}|/E_{T}^{\mu^{+}\mu^{-}} <$ 0.4 \\
  &$\text{cos}(\text{angle}_{3D})$ $< -0.75$ \\ 
  &$N_{jets} < 1$ \\
   \hline
    \end{tabular}
    \label{table:selection2}
\end{table}
We present the distribution of cos$\theta_{CS}$ for a resonant model based on Einstein-Cartan theory in Figure \ref{fig2}. The model assumes a dark boson ($A^{\prime} \rightarrow \mu^{+}\mu^{-}$) with mass ($M_{A'} =$ 500 GeV), and we compare it with Drell-Yan events. All events meet the pre-selection criteria outlined in Table \ref{table:selection2} and have a reconstructed invariant mass ranging from 460 to 540 GeV. The results are illustrated with black open circles representing the model signal and blue closed circles for the Drell-Yan events, normalized to unity. 
While in Figure \ref{figure-spin}, the distribution of cos$\theta_{CS}$ is presented for a resonant model with a mass of \( A^{\prime} = 500 \, \text{GeV} \) at a center-of-mass energy of \( \sqrt{s} = 14 \, \text{TeV} \) and an integrated luminosity of \( 3000 \, \text{fb}^{-1} \). 
The Monte Carlo (MC) data for the EC model are fitted using the function described in \cite{cms-note} 
\begin{equation}
\text{par[0]} \big(1 - \text{cos}\theta^{4}_{CS} \big),
\end{equation}
for graviton decay into dilepton ($G^* \rightarrow f\bar{f}$), 
represented by the red curve.

We observe a clear distinction between the simplified model and the Drell-Yan events. The signal shape exhibits a typical characteristic of a spin-2 boson, displaying a symmetric distribution around zero. This distribution aligns with the findings from the study conducted in \cite{Osland, spin, cms-note}.
\begin{figure*}
\centering
\subfigure[]{
  \includegraphics[width=70mm]{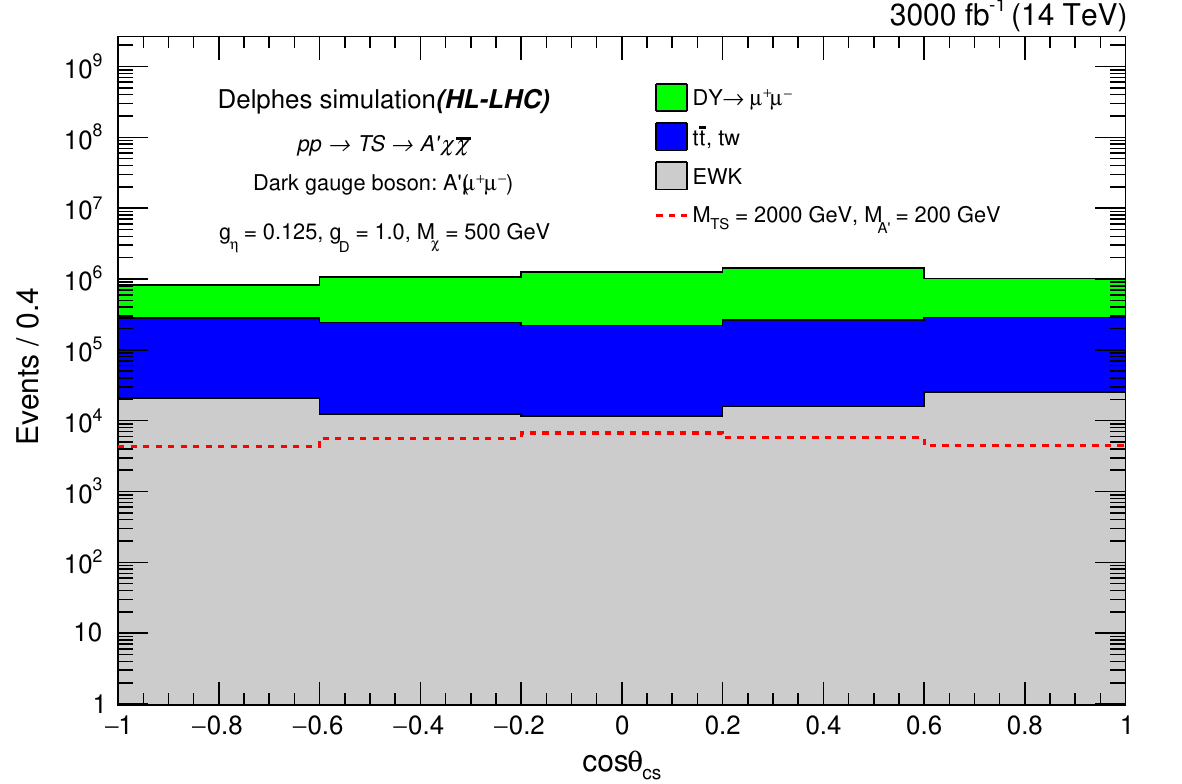}
  \label{bin1}
}
\hspace{0mm}
\subfigure[]{
  \includegraphics[width=70mm]{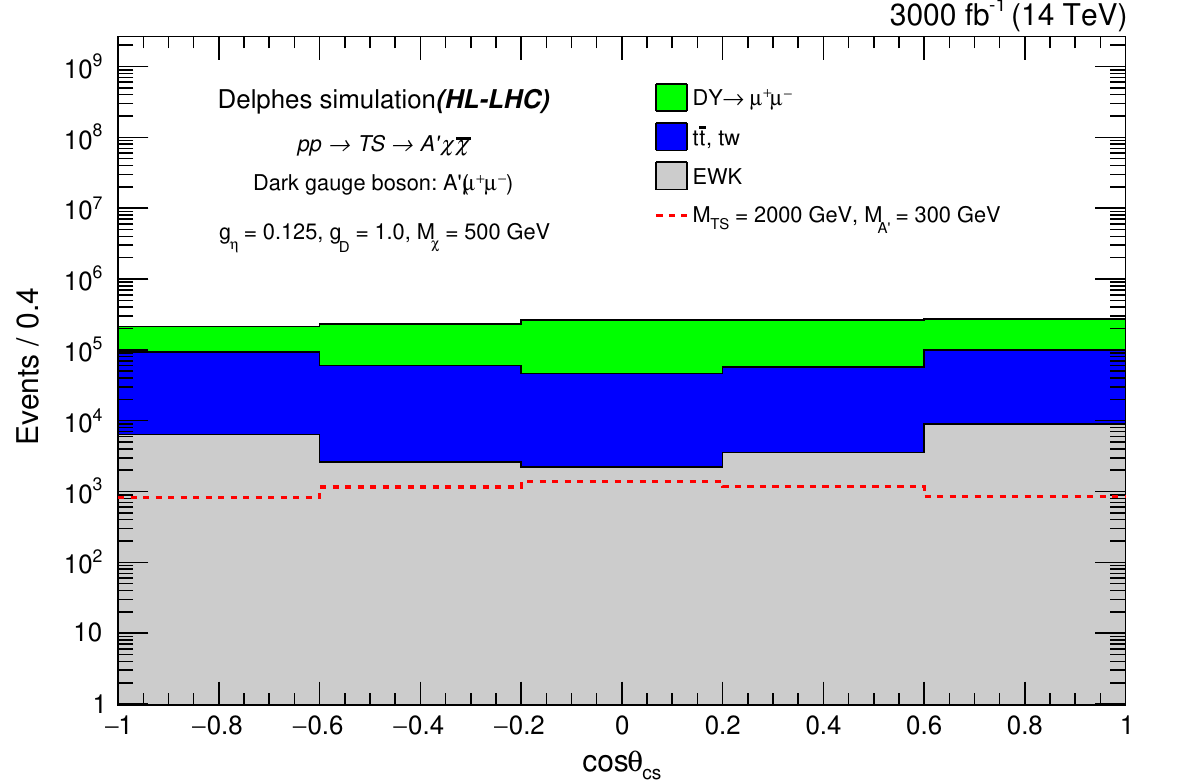}
  \label{bin2}
}
\hspace{0mm}
\subfigure[]{
  \includegraphics[width=70mm]{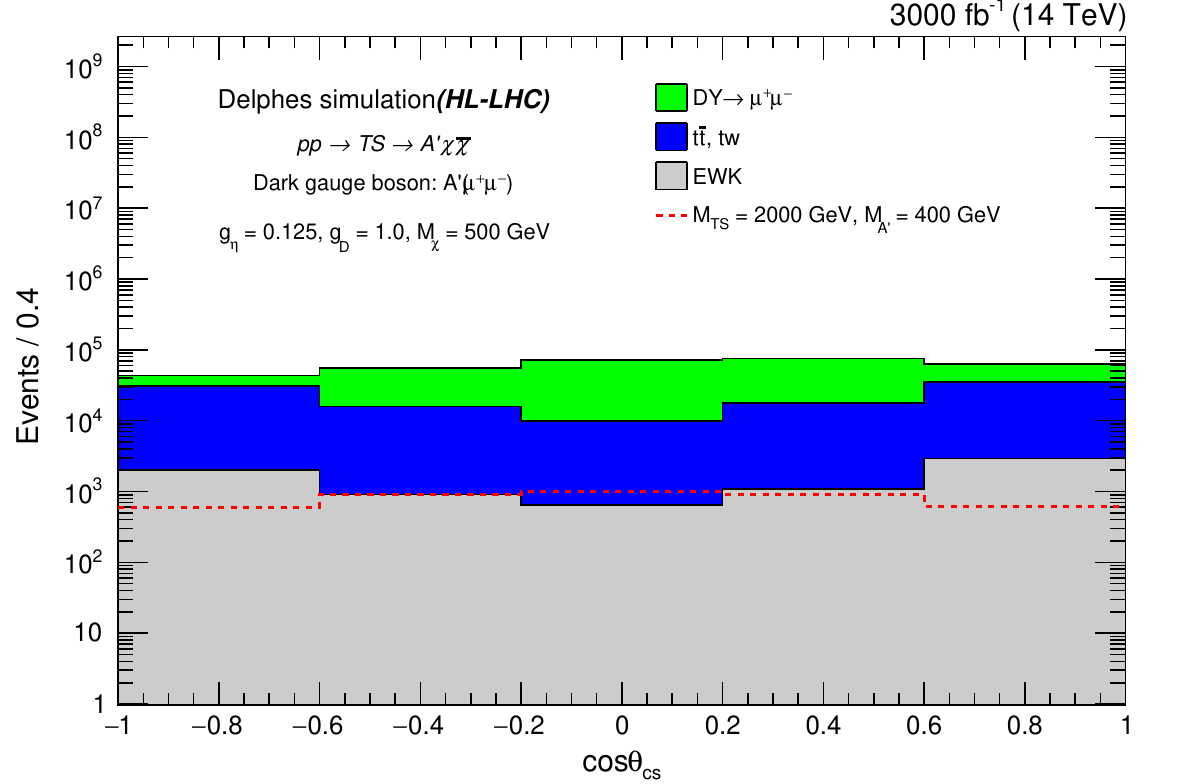}
  \label{bin3}
}
\hspace{0mm}
\subfigure[]{
  \includegraphics[width=70mm]{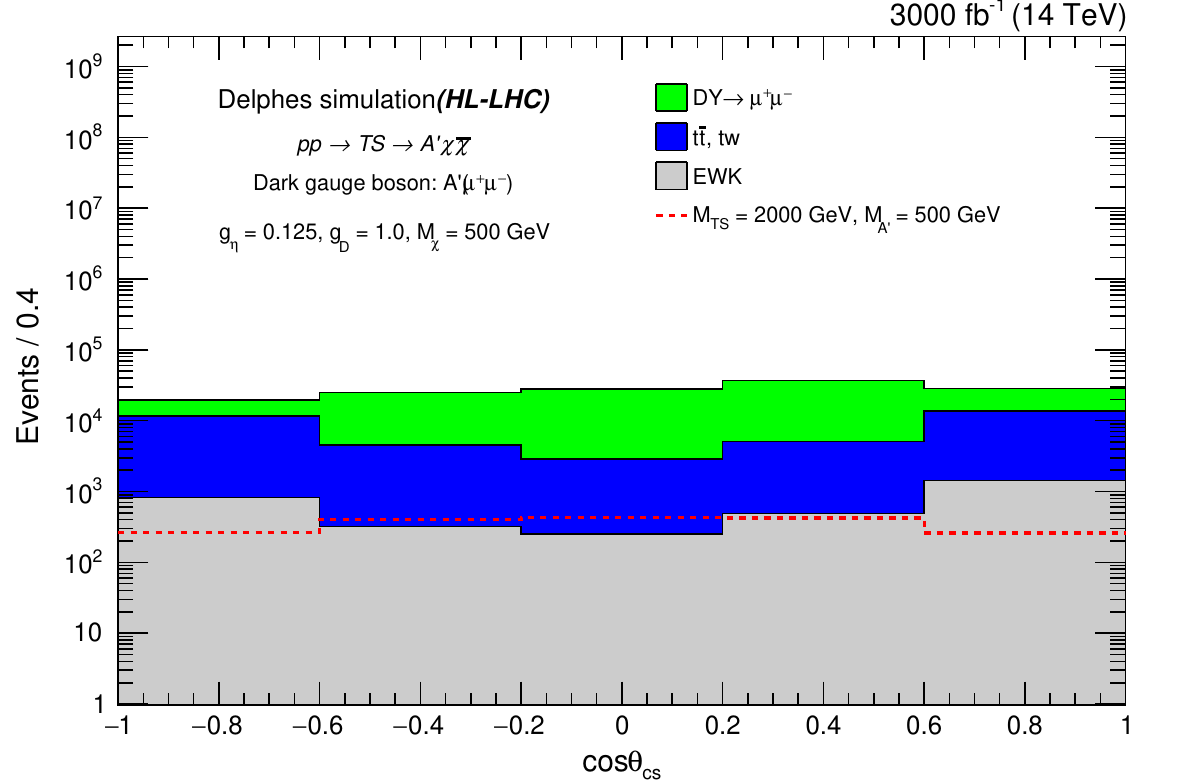}
  \label{bin4}
}
\hspace{0mm}
\subfigure[]{
  \includegraphics[width=70mm]{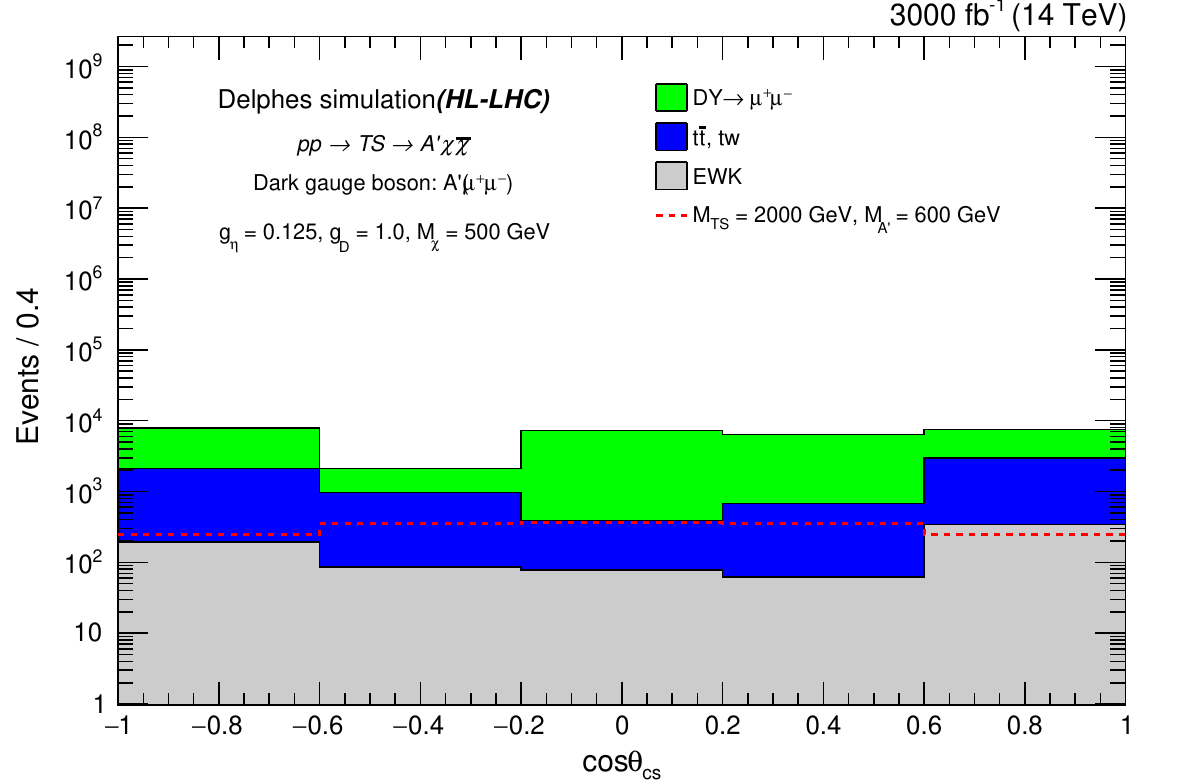}
  \label{bin5}
}
\caption{Distributions of cos$\theta_{CS}$ are illustrated, for events passing pre-selection listed in table \ref{table:selection2}, for the standard model expectations (histograms) for several dimuon mass windows: 
160 $< M_{\mu^+\mu^-} <$ 240 GeV \ref{bin1}, 
260 $< M_{\mu^+\mu^-} <$ 340 GeV \ref{bin2}, 
360 $< M_{\mu^+\mu^-} <$ 440 GeV \ref{bin3}, 
460 $< M_{\mu^+\mu^-} <$ 540 GeV \ref{bin4}, and
560 $< M_{\mu^+\mu^-} <$ 640 GeV \ref{bin5}. 
The signal presentation of the model corresponding to the Einstein-Cartan theory with the value of $M_{A^{\prime}}$ runs from 200 to 600 GeV is superimposed.}
\label{fig3}
\end{figure*}

The graphs in figure \ref{fig3} show the distribution of cos$\theta_{CS}$ across different mass bins: 
160 $< M_{\mu^+\mu^-} <$ 240 GeV \ref{bin1}, 
260 $< M_{\mu^+\mu^-} <$ 340 GeV \ref{bin2}, 
360 $< M_{\mu^+\mu^-} <$ 440 GeV \ref{bin3}, 
460 $< M_{\mu^+\mu^-} <$ 540 GeV \ref{bin4}, and
560 $< M_{\mu^+\mu^-} <$ 640 GeV \ref{bin5}. 
The histograms in green, blue, and gray represent Drell-Yan, $t\bar{t}$ + tW, and vector boson pair backgrounds, respectively, stacked together, with the red dotted line indicating the Einstein-Cartan signal for the dark neutral gauge boson A$^{\prime}$, with $M_{\chi} = 500$ GeV and $M_{TS} = 2000$ GeV. 

\section{Results}
\label{section:Results}
After applying the pre-selection cuts outlined in Table \ref{table:selection2}, the Einstein-Cartan signal events were completely obscured by the SM background. It was determined that the signal and background events could not be distinguished solely based on the distribution of the cos$\theta_{CS}$ variable. Therefore, a tighter selection criterion was optimized to effectively differentiate the model signals from the SM backgrounds. 

The tight selection is based on five variables: 

1. We calculate the azimuthal angle difference \( \Delta\phi_{\mu^{+}\mu^{-},\vec{E}^{\text{miss}}_{T}} \), which represents the difference between the azimuthal angles of the dimuon and the missing transverse energy \( (|\phi^{\mu^{+}\mu^{-}} - \phi^{\text{miss}}|) \). This value must be greater than 2.5 radians.

2. We evaluate the relative difference between the transverse energy of the dimuon \( (E_{T}^{\mu^{+}\mu^{-}}) \) and the missing transverse energy \( (E^{\text{miss}}_{T}) \). This difference is set to be less than 0.4, defined by the condition \( |E_{T}^{\mu^{+}\mu^{-}} - E^{\text{miss}}_{T}|/E_{T}^{\mu^{+}\mu^{-}} < 0.4 \). 

3. We impose a constraint on the cosine of the 3D angle between the missing energy vector and the dimuon system vector to ensure they are oriented back-to-back, requiring that \( \cos(\text{angle}_{3D}) < -0.75 \).

4. The number of jets ($N_{jets}$) with $p^{j}_{T} > 20$ GeV and $|\eta^{j}| < 2.5$ should be less than 1.

Lastly, we limit the invariant mass of the dimuon to a range centered around the mass of the neutral gauge boson \( A^{\prime} \). Specifically, we require that \( M_{A^{\prime}} - 40 < M_{\mu^{+}\mu^{-}} < M_{A^{\prime}} + 40 \).

\begin{figure}
\centering
\subfigure[]{
  \includegraphics[width=70.mm]{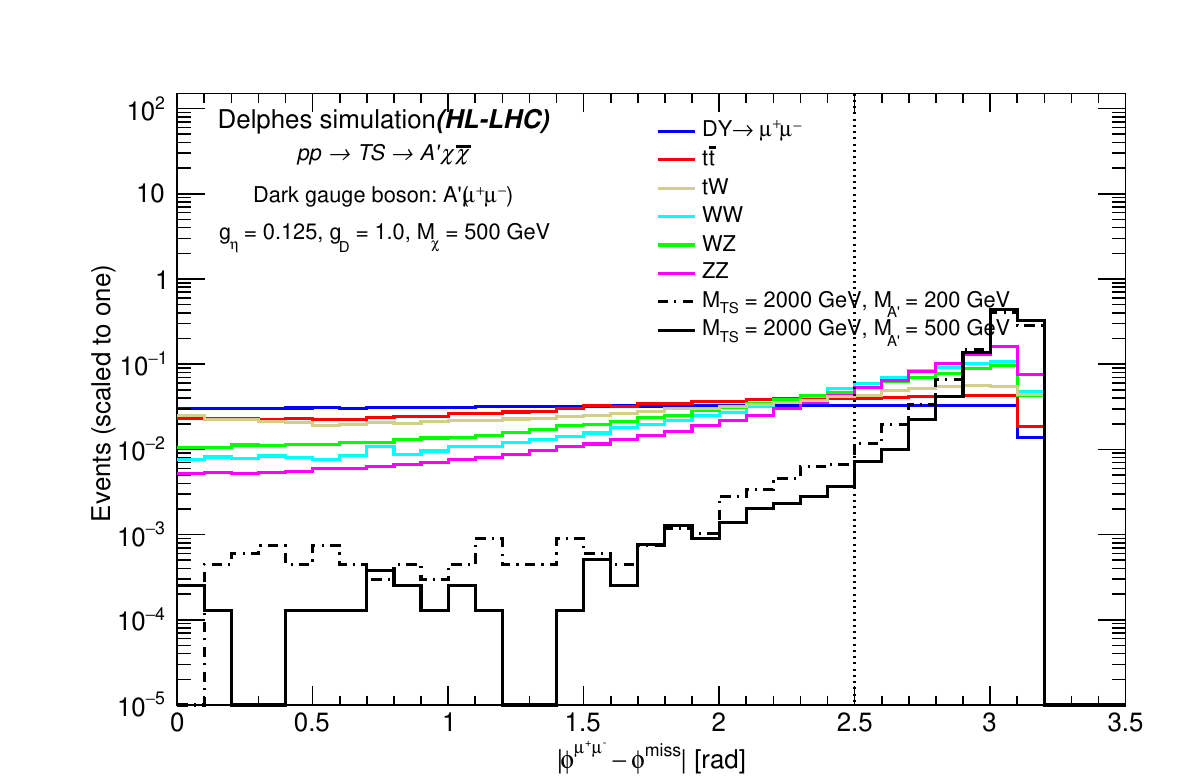} 
  \label{deltaphi}
}
\subfigure[]{
  \includegraphics[width=70.0mm]{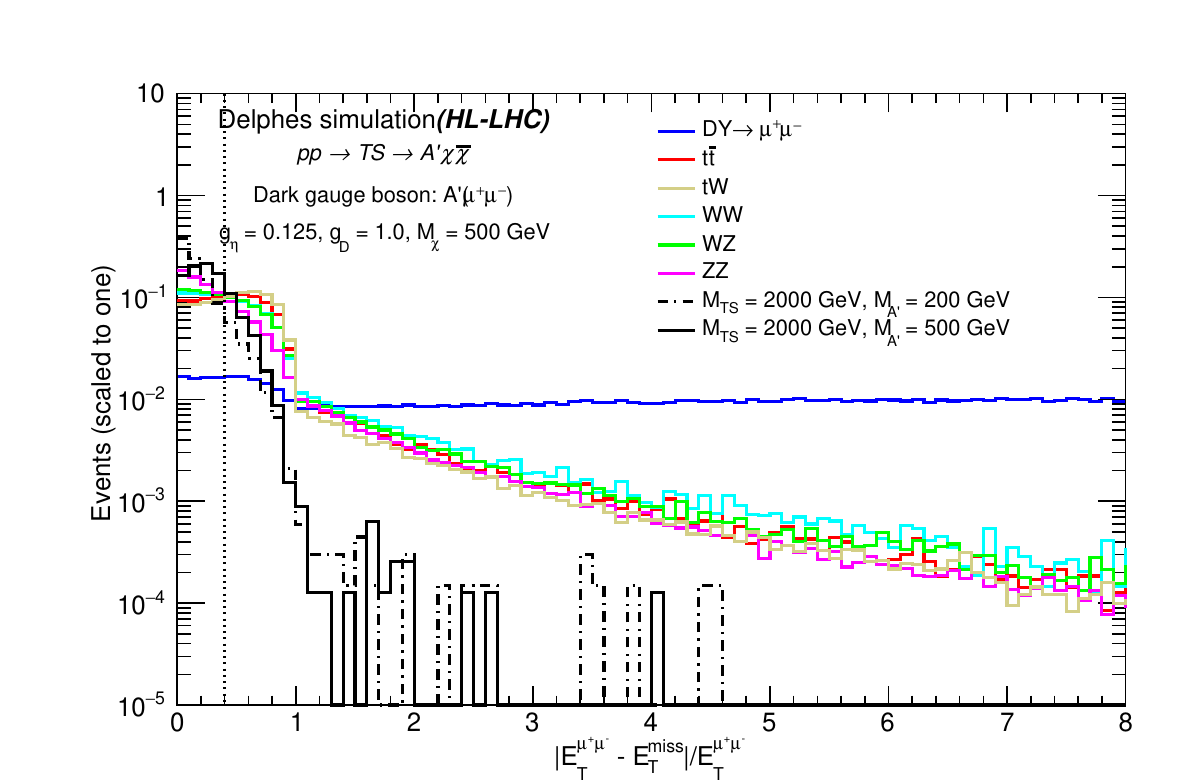}
  \label{ptdiff}
}
\hspace{0mm}
\subfigure[]{
  \includegraphics[width=70.0mm]{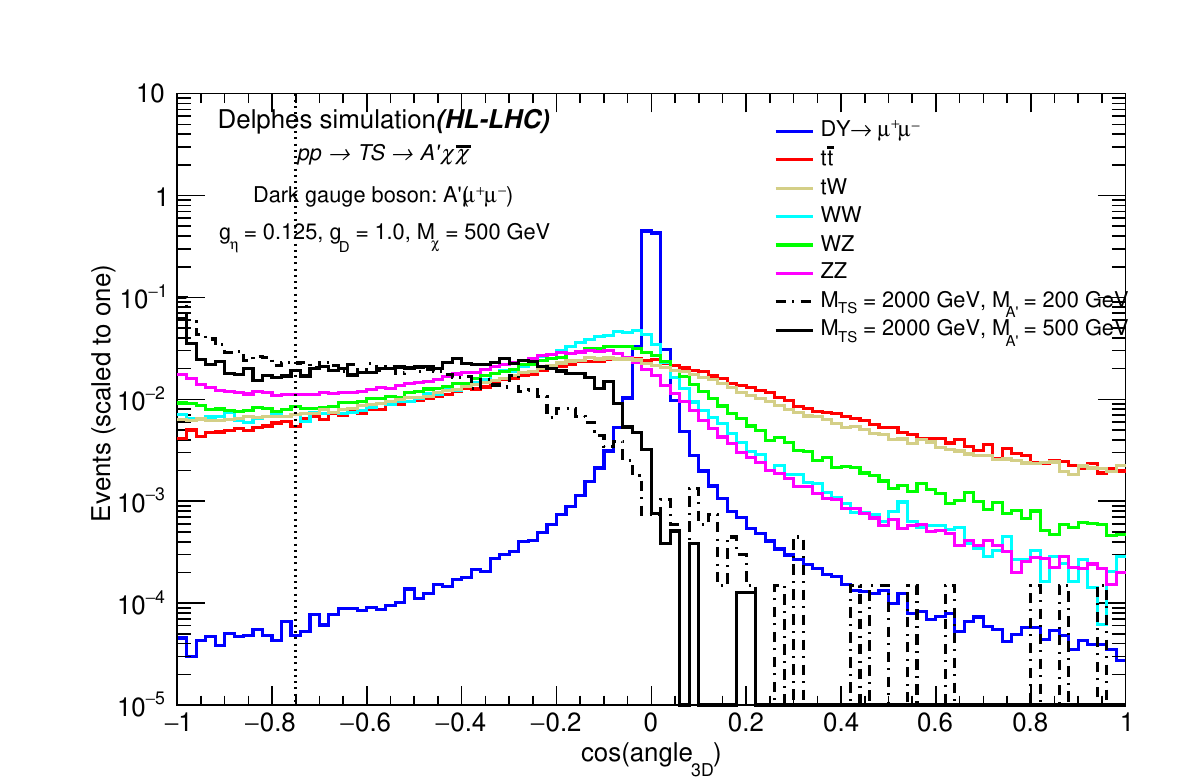}
  \label{3Dangle}
}
\hspace{0mm}
\subfigure[]{
  \includegraphics[width=70.0mm]{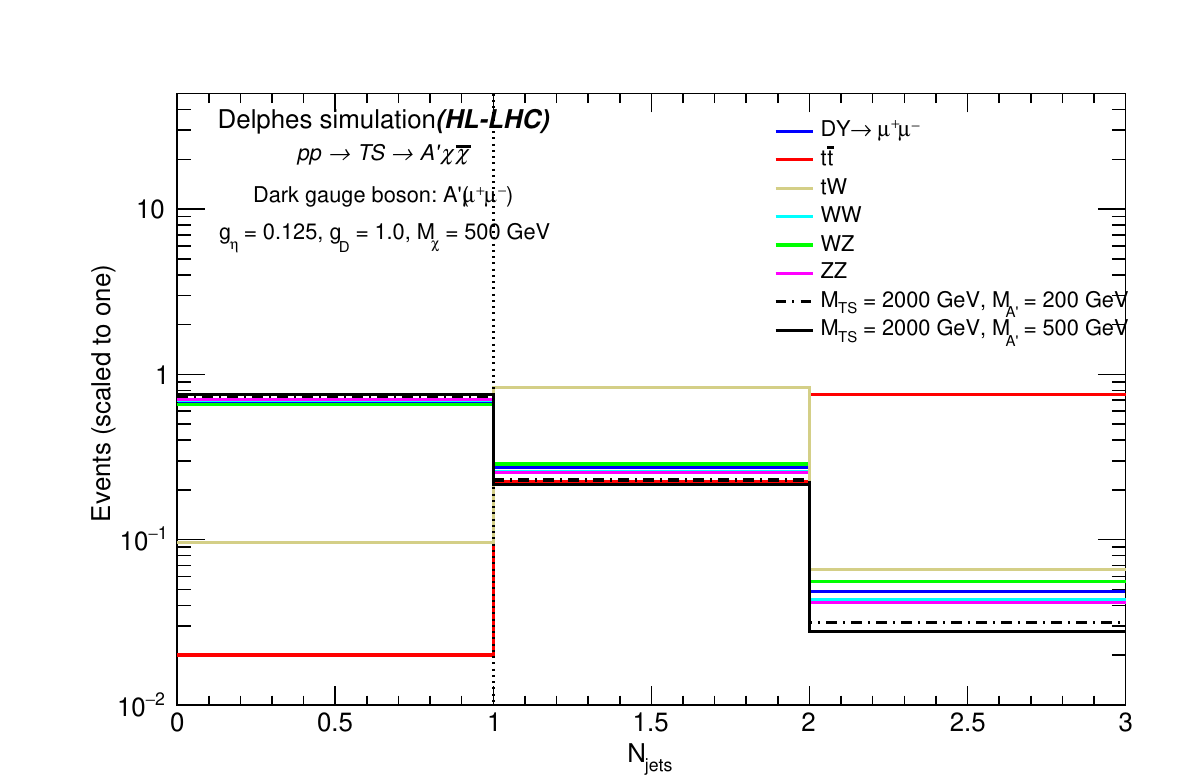}
  \label{Njets}
}
\caption{Distributions of $\Delta\phi_{\mu^{+}\mu^{-},\vec{E}_{T}^{\text{miss}}}$
\ref{deltaphi}, 
$|E_{T}^{\mu^{+}\mu^{-}} - E_{T}^{\text{miss}}|/E_{T}^{\mu^{+}\mu^{-}}$ \ref{ptdiff}, $\text{cos}(\text{angle}_{3D})$ \ref{3Dangle}, and the number of jets \ref{Njets}
for the signal presentations of the model corresponding to the Einstein-Cartan theory with $M_{A^{\prime}} =$ 200 and 500 GeV and SM backgrounds, for dimuon events with each muon passing the pre-selection summarized in table \ref{table:selection2}.} 
\label{cuts}
\end{figure}

In Figure \ref{cuts}, we present the distributions of several variables for dimuon events, where each muon meets the pre-selection criteria outlined in Section \ref{section:AnSelection}. These variables include $\Delta\phi_{\mu^{+}\mu^{-},\vec{E}_{T}^{\text{miss}}}$ \ref{deltaphi}, the normalized difference $|E_{T}^{\mu^{+}\mu^{-}} - E_{T}^{\text{miss}}|/E_{T}^{\mu^{+}\mu^{-}}$ \ref{ptdiff}, $\text{cos}(\text{angle}_{3D})$ \ref{3Dangle}, and the number of jets ($N_{jets}$) \ref{Njets}. 
The dotted and solid black histograms illustrate two signal scenarios from the Einstein-Cartan theory model, corresponding to \(M_{A^{\prime}}\) values of 200 GeV and 500 GeV. These scenarios feature a dark matter mass of \(M_{\chi} = 500\) GeV and mass of the torsion field $M_{TS} = 2000$ GeV, with coupling constants \(\texttt{g}_{\eta} = 0.125\) and \(\texttt{g}_{D} = 1.0\). Additionally, the Standard Model backgrounds are represented by colored histograms.
For clarity, all distributions have been normalized to one, and the vertical dashed lines indicate the selected cut values for each variable.

Table \ref{table:selection2} summarizes the strict selection criteria. By applying tighter cuts, in addition to those specified in the pre-selection, we eliminated the DY and ZZ backgrounds. Furthermore, there was a significant reduction in the contributions from the \(t\bar{t}\), tW, WW, and WZ backgrounds.

\begin{figure}
\centering
\subfigure[N-1 Eff. for $\Delta\phi_{\mu^{+}\mu^{-},\vec{E}_{T}^{\text{miss}}} >$ 2.5]{
  \includegraphics[width=70.mm]{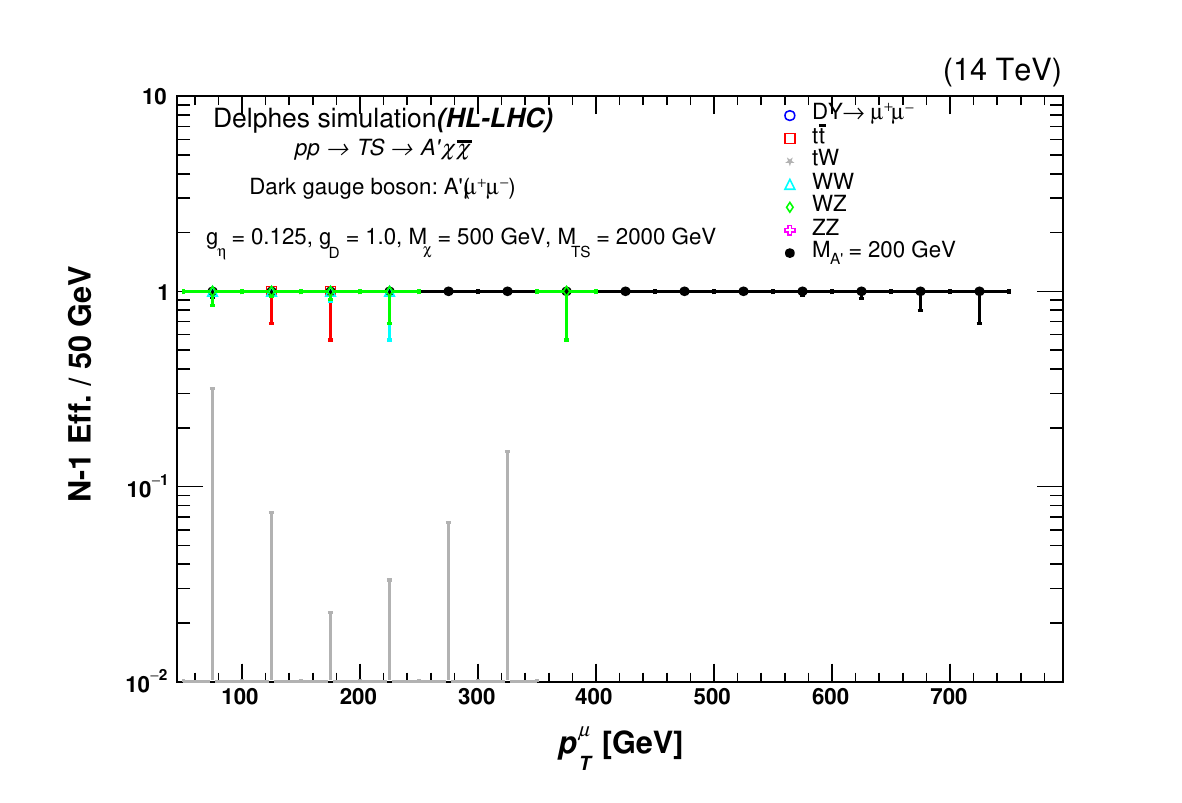} 
  \label{Effdeltaphi}
}
\subfigure[N-1 Eff. for $|E_{T}^{\mu^{+}\mu^{-}}- E_{T}^{\text{miss}}|/E_{T}^{\mu^{+}\mu^{-}} <$ 0.4]{
  \includegraphics[width=70.0mm]{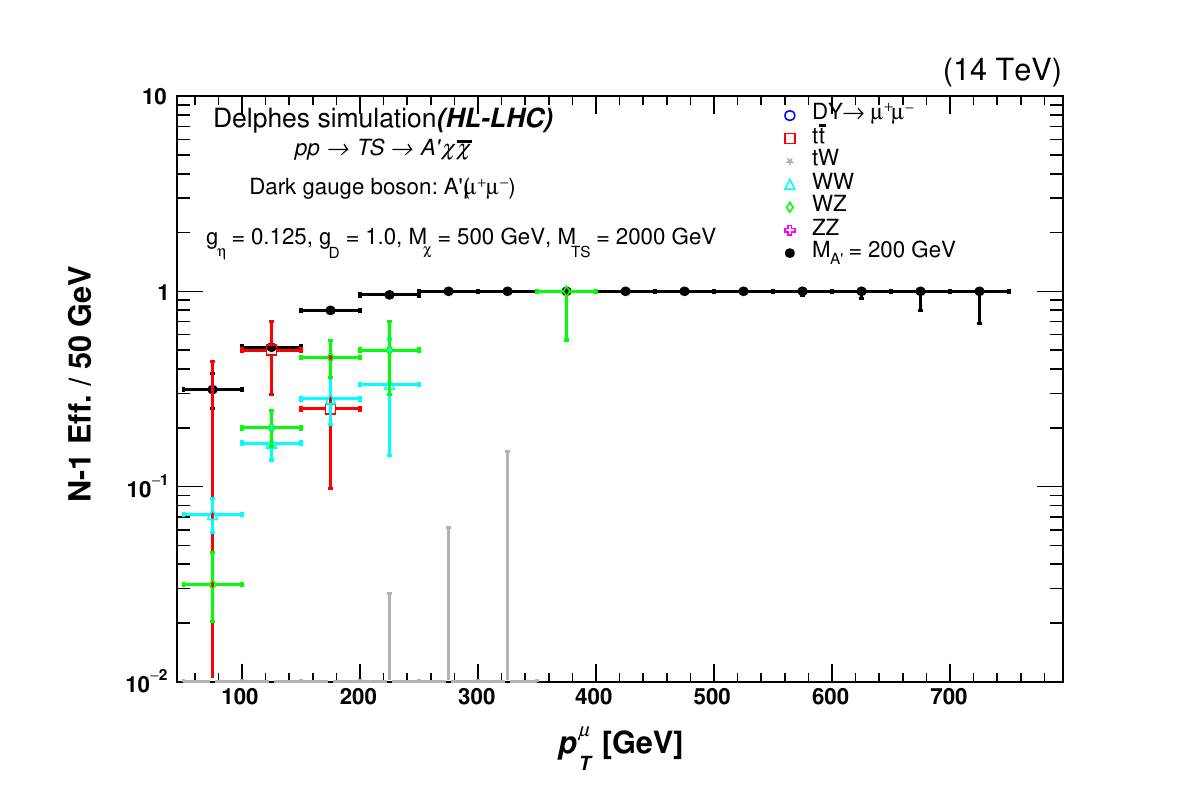}
  \label{Effptdiff}
}
\hspace{0mm}
\subfigure[N-1 Eff. for $\text{cos}(\text{angle}_{3D})$ $< -0.75$]{
  \includegraphics[width=70.0mm]{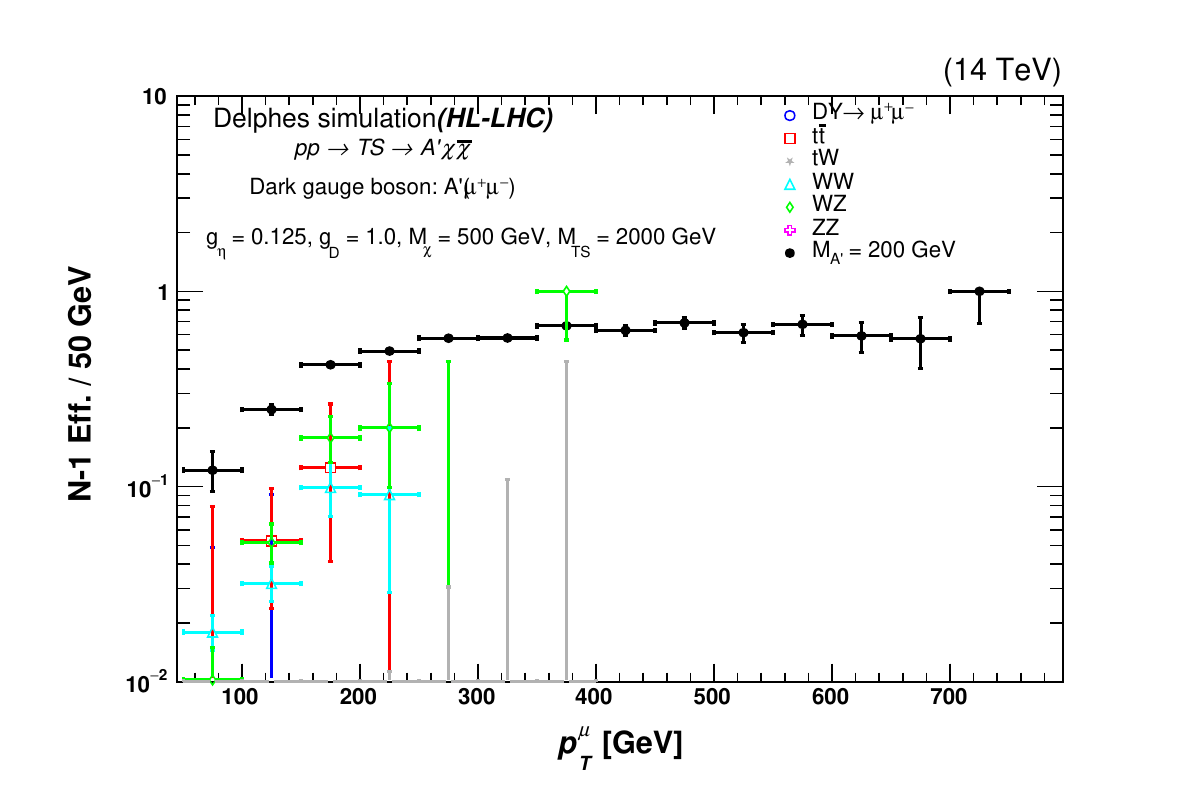}
  \label{Eff3Dangle}
}
\hspace{0mm}
\subfigure[N-1 Eff. for $N_{jets} < 1$]{
  \includegraphics[width=70.0mm]{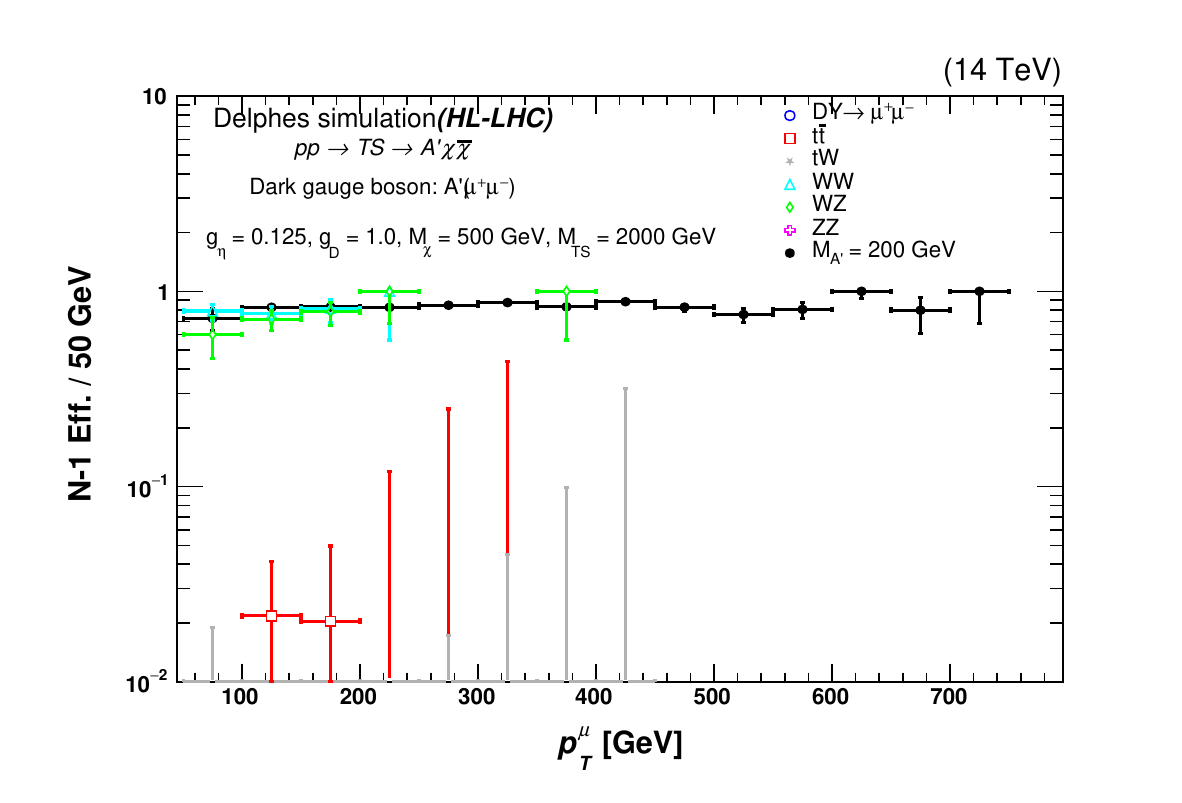}
  \label{EffNjets}
}
\caption{Distributions of the N-1 efficiencies plotted against the transverse momentum of the leading reconstructed muon ($p^{\mu}_{T}$) for the 4 analysis cuts.
Illustrated for the signal in the EC gravity scenario with $M_{TS} = 2000$ GeV, $M_{A'} = 200$ GeV, and coupling constants $\texttt{g}_{\eta} = 0.125$ and $\texttt{g}_{D} = 1.0$ and
for the SM backgrounds.} 
\label{Effs}
\end{figure}
The figures of merit for optimizing these stringent criteria are represented by plotting the N-1 efficiency for each of the four criteria mentioned earlier. The N-1 efficiency is calculated by dividing the number of events that successfully pass the final selection, as shown in Table \ref{table:selection2}, by the number of events that would pass the final selection if the specific cut being considered were not applied.

In Figure \ref{Effs}, we present the distributions of the N-1 efficiencies plotted against the transverse momentum of the leading reconstructed muon (\( p^{\mu}_{T} \)) within the mass window ($160 < M_{\mu^{+}\mu^{-}} < 240$). The following cuts have been studied: 
$\Delta\phi_{\mu^{+}\mu^{-},\vec{E}_{T}^{\text{miss}}} > 2.5$ \ref{Effdeltaphi},
$|E_{T}^{\mu^{+}\mu^{-}} - E_{T}^{\text{miss}}| / E_{T}^{\mu^{+}\mu^{-}} < 0.4$ \ref{Effptdiff},
$\text{cos}(\text{angle}_{3D}) < -0.75$ \ref{Eff3Dangle}, and 
$N_{jets} < 1$ \ref{EffNjets}. 
These plots focus on the signal in the EC gravity scenario (indicated by black closed circles), with $M_{TS} = 2000$ GeV, $M_{A'} = 200$ GeV, and coupling constants $\texttt{g}_{\eta} = 0.125$ and $\texttt{g}_{D} = 1.0$, alongside standard model (SM) backgrounds marked with open colored markers.

The efficiency plots indicate that implementing these four stringent selection cuts significantly suppresses backgrounds from DY and ZZ processes, while also minimizing contamination from $t\bar{t}$, tW, WW, and WZ events. Furthermore, this approach guarantees that the signal maintains a consistently flat efficiency at high muons 
$p_T$ ($p^{\mu}_T > 200$ GeV).

\begin{figure*}
\centering
\subfigure[]{
  \includegraphics[width=70mm]{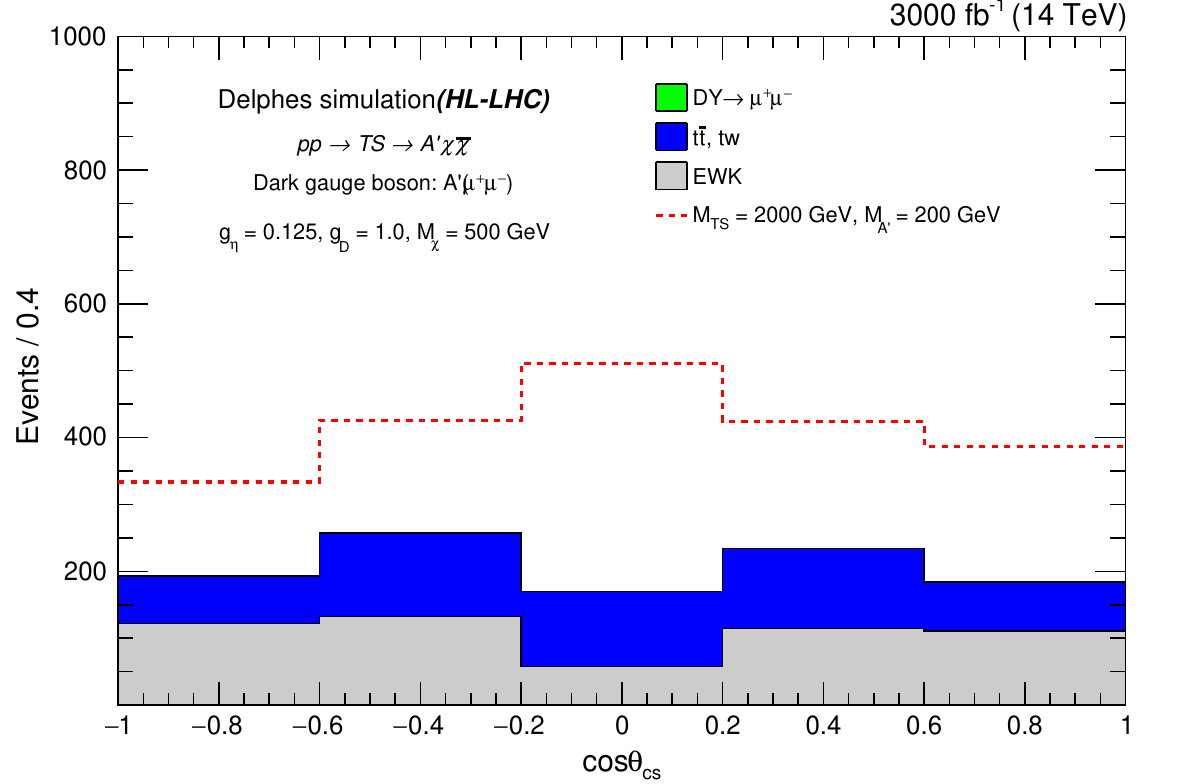}
  \label{bin200}
}
\subfigure[]{
  \includegraphics[width=70mm]{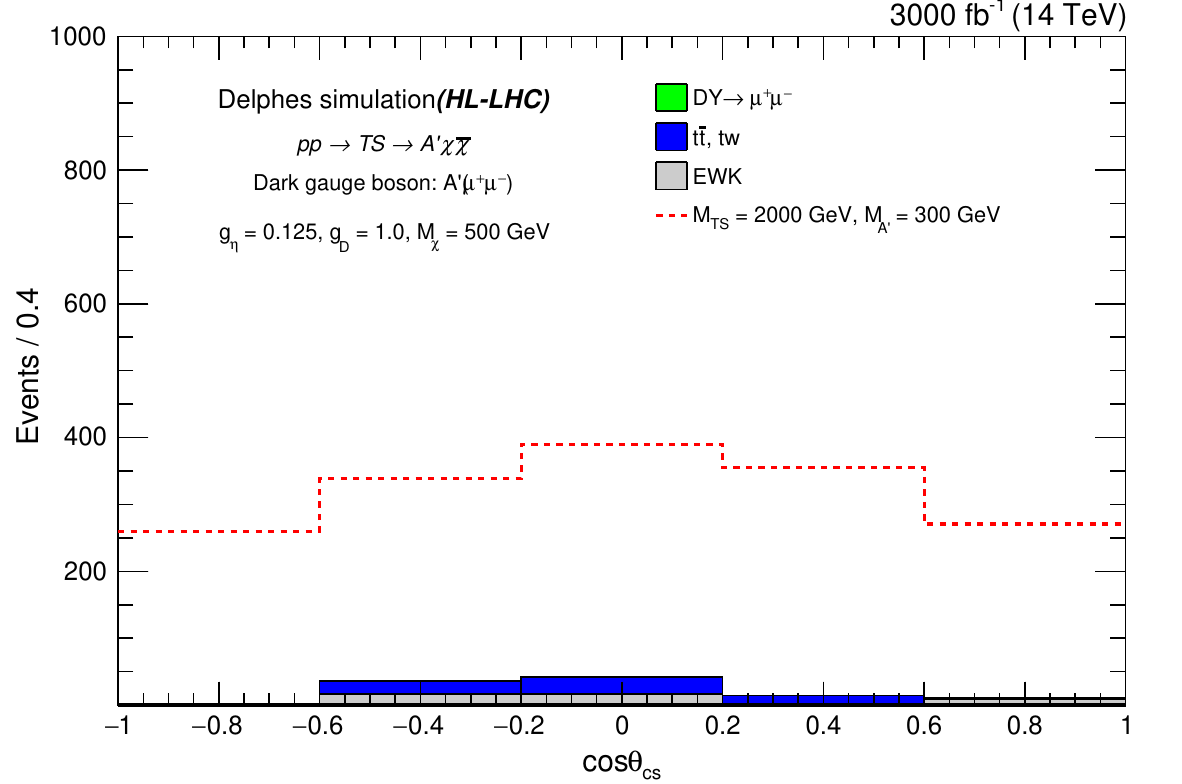}
  \label{bin300}
}
\hspace{0mm}
\subfigure[]{
  \includegraphics[width=70mm]{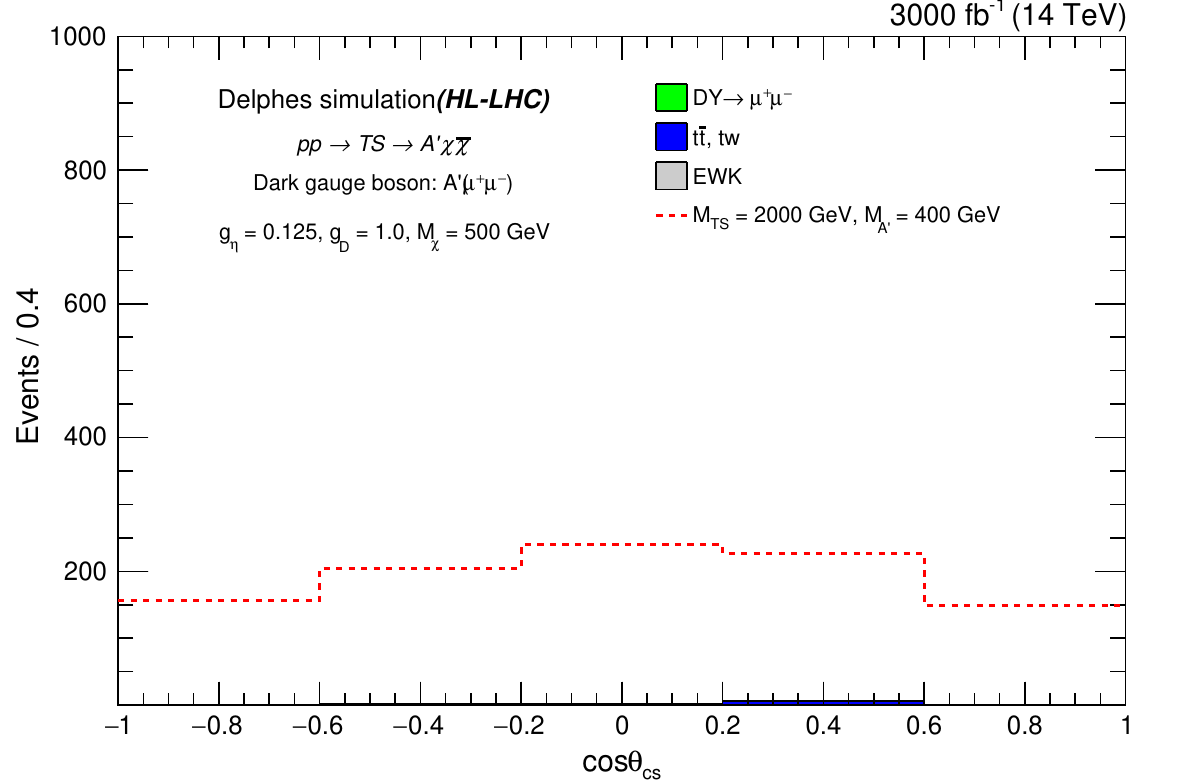}
  \label{bin400}
}
\subfigure[]{
  \includegraphics[width=70mm]{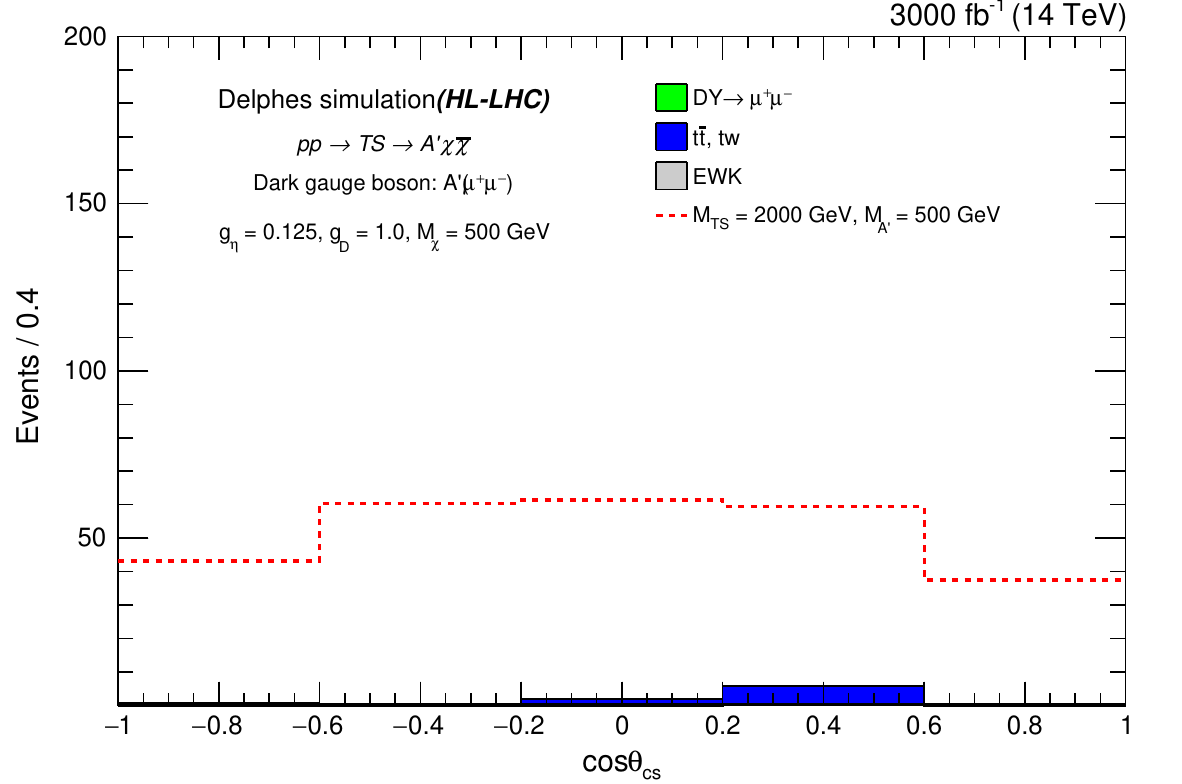}
  \label{bin500}
}
\hspace{0mm}
\subfigure[]{
  \includegraphics[width=70mm]{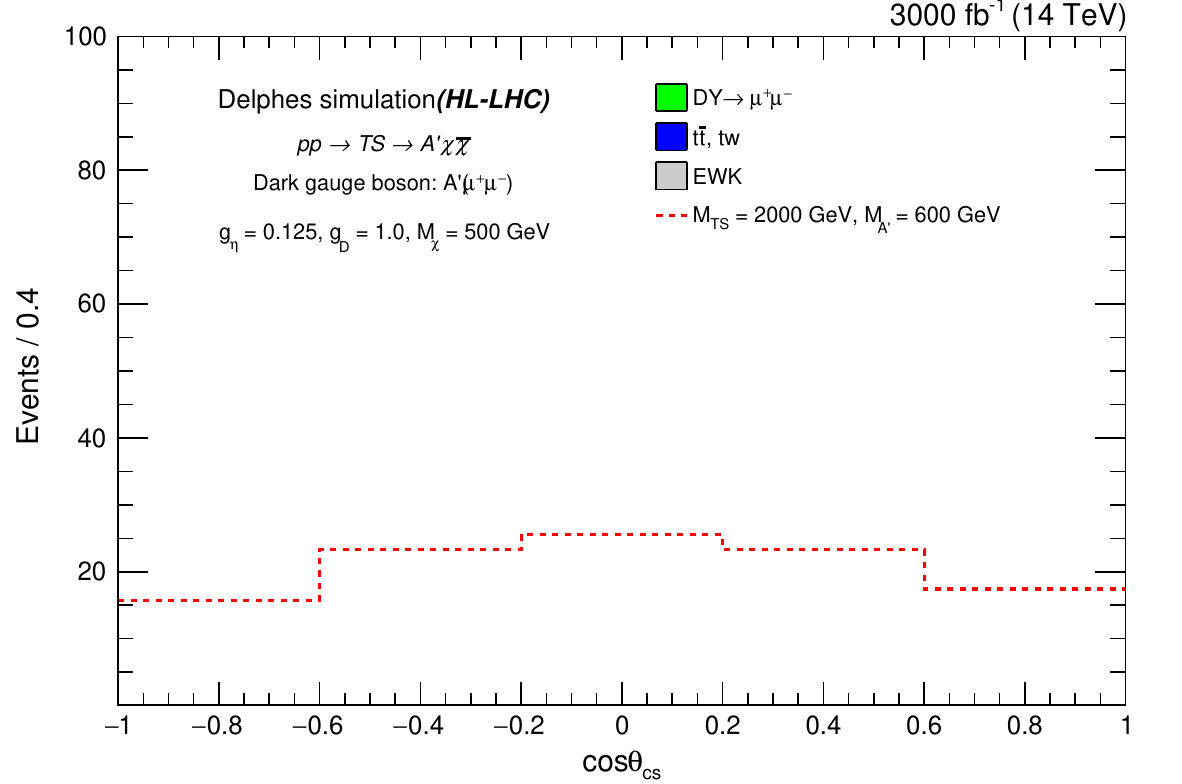}
  \label{bin600}
}
\hspace{0mm}
\caption{Distributions of cos$\theta_{CS}$ are illustrated, for events passing final selection listed in table \ref{table:selection2}, for the standard model expectations (histograms) for several dimuon mass windows: 
160 $< M_{\mu^+\mu^-} <$ 240 GeV \ref{bin200}, 
260 $< M_{\mu^+\mu^-} <$ 340 GeV \ref{bin300}, 
360 $< M_{\mu^+\mu^-} <$ 440 GeV \ref{bin400}, 
460 $< M_{\mu^+\mu^-} <$ 540 GeV \ref{bin500}, and
560 $< M_{\mu^+\mu^-} <$ 640 GeV \ref{bin600}. 
The signal presentation of the model corresponding to the Einstein-Cartan theory with the value of $M_{A^{\prime}}$ runs from 200 to 600 GeV is superimposed.}
\label{fig8}
\end{figure*}

The graphs depicted in Figure \ref{fig8} show the distributions of the cos$\theta_{CS}$ for events that have passed the event final selection criteria mentioned in Tables \ref{table:selection2} in multiple mass bins:
160 $< M_{\mu^+\mu^-} <$ 240 GeV \ref{bin200}, 
260 $< M_{\mu^+\mu^-} <$ 340 GeV \ref{bin300}, 
360 $< M_{\mu^+\mu^-} <$ 440 GeV \ref{bin400}, 
460 $< M_{\mu^+\mu^-} <$ 540 GeV \ref{bin500}, and
560 $< M_{\mu^+\mu^-} <$ 640 GeV \ref{bin600}. 
Applying the final cuts mentioned in table \ref{table:selection2} demonstrates that the signal samples are easily distinguishable from the SM backgrounds.

To determine the required values of \(M_{A^{\prime}}\) necessary for observing either a deviation or a potential discovery, we have assessed the signal significance, defined as \(\it{S} = N_s/\sqrt{N_s + N_b}\), by varying \(M_{A^{\prime}}\). In this formula, \(N_s\) represents the number of signal events, while \(N_b\) denotes the total number of Standard Model background events that pass the final selection criteria outlined in Table \ref{table:selection2}.

\begin{figure}
\centering
\subfigure[]{
  \includegraphics[width=70.0mm]{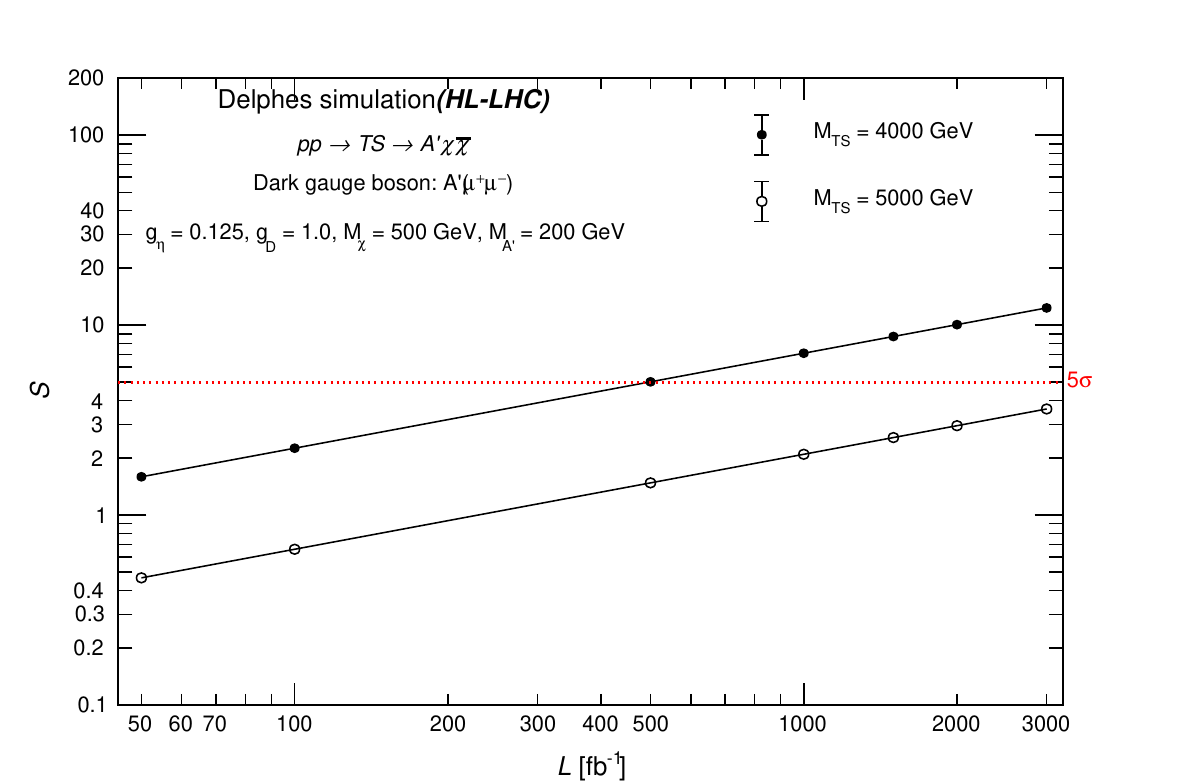}
  \label{figure:s1}
}
\hspace{0mm}
\centering
\subfigure[]{
  \includegraphics[width=70.0mm]{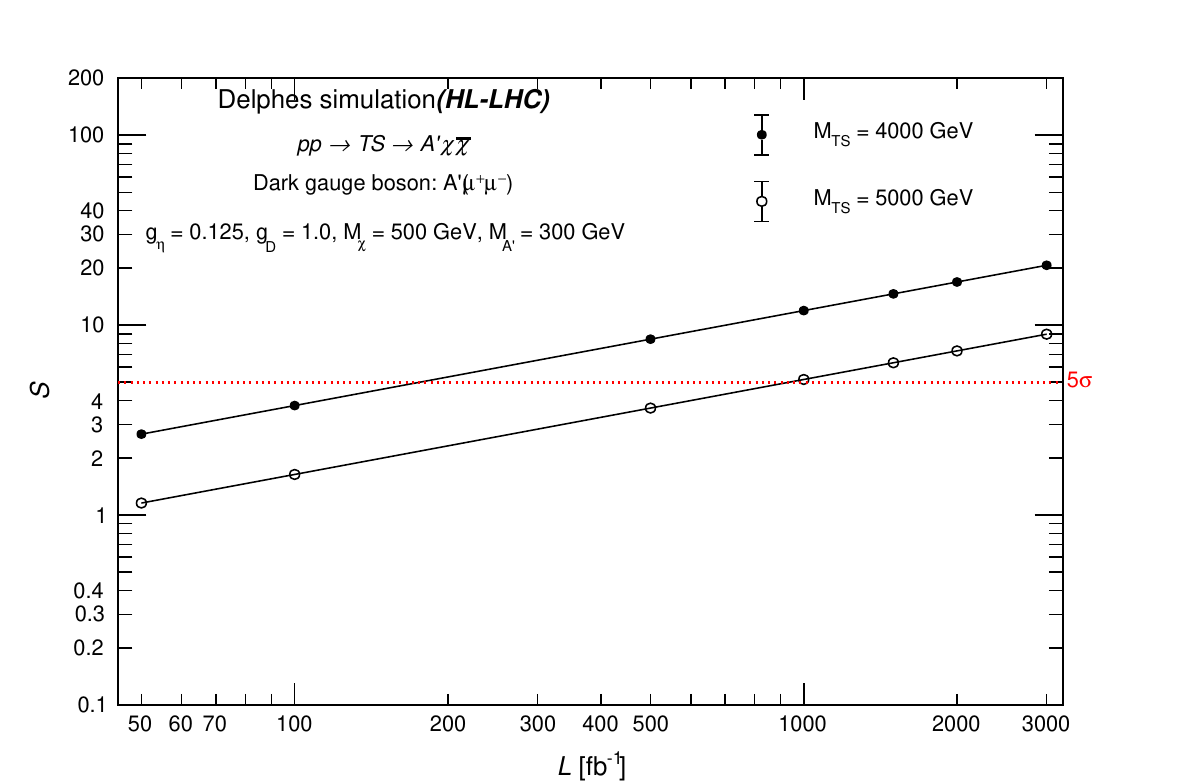}
  \label{figure:s2}
}
\hspace{0mm}
\centering
\subfigure[]{
  \includegraphics[width=70.0mm]{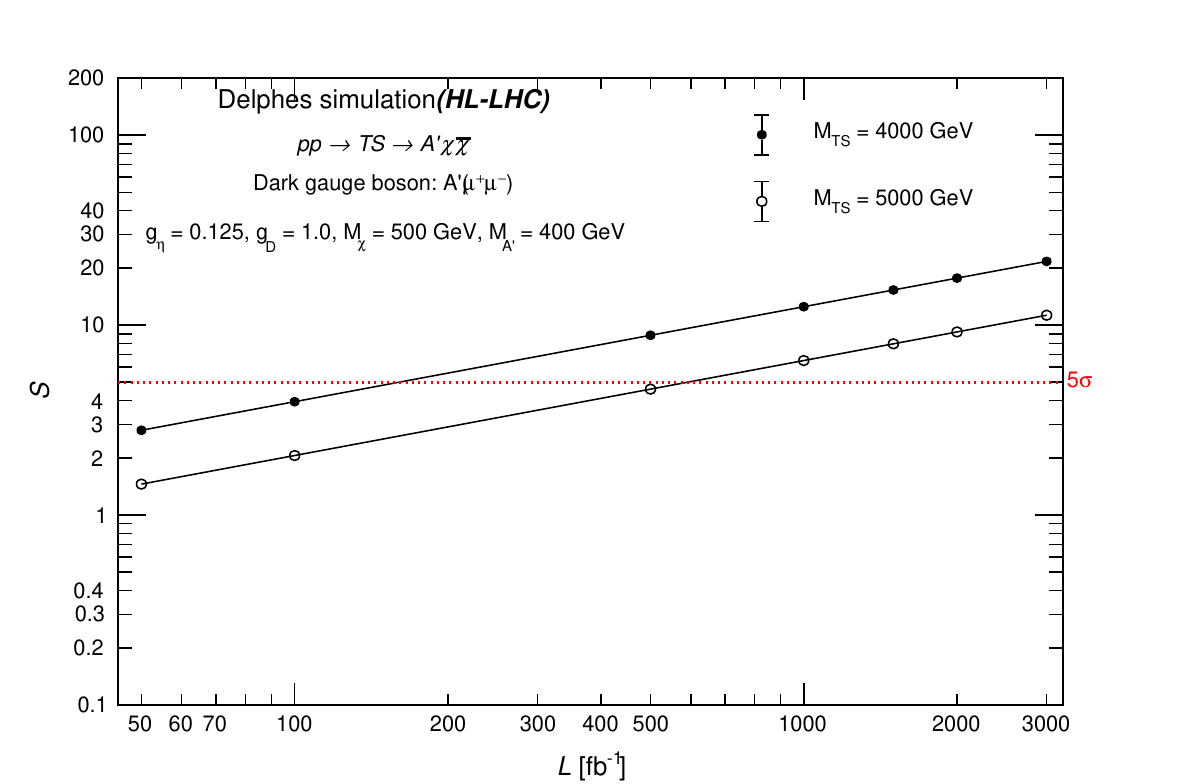}
  \label{figure:s3}
}
\caption{The significance (\(\it{S}\)) is plotted against the integrated luminosity ($L$) for events that pass the full set of cuts listed in Table \ref{table:selection2}. This is shown for different tensor scalar masses (\(M_{TS}\)): \(M_{A^{\prime}} = 200\) GeV in plot \ref{figure:s1}, 300 GeV in plot \ref{figure:s2}, and 400 GeV in plot \ref{figure:s3}.  
The model relates to Einstein-Cartan theory with coupling constants $\texttt{g}_{\eta} = 0.125$ and $\texttt{g}_{D} = 1.0$, assuming the dark matter mass is $M_{\chi} = 500$ GeV. The dashed horizontal red line indicates $\it{S} = 5$.}
\label{figure:significance2}
\end{figure}

The plots illustrated in Figure \ref{figure:significance2} display the connection between signal significance ($\it{S}$) and integrated luminosity($\it{L}$) for a specific dark matter mass scenario, namely \(M_{\chi} = 500\) GeV, using coupling constants of \(\texttt{g}_{\eta} = 0.125\) and \(\texttt{g}_{D} = 1.0\). Plot \ref{figure:s1} centers on \(M_{A^{\prime}} = 200\) GeV, plot \ref{figure:s2} focuses on \(M_{A^{\prime}} = 300\) GeV, and plot \ref{figure:s3} examines \(M_{A^{\prime}} = 400\) GeV, all for events that satisfy the final criteria outlined in Table \ref{table:selection2}. The dashed red line in these plots indicates a significance value of \(\it{S}=5\).

In Table \ref{tab:5_sigma}, we present the calculations for the required luminosity to achieve 5$\sigma$ significance, denoted as \(\mathcal{L}_{5\sigma}^{\text{Req}}\). This is evaluated for two values of \(M_{TS}\), specifically 4000 GeV and 5000 GeV, against the Standard Model background. The analysis spans three different values of \(M_{A^{\prime}}\) and utilizes coupling constants of \(\texttt{g}_{\eta} = 0.125\) and \(\texttt{g}_{D} = 1.0\), with \(M_{\chi} = 500\) GeV, all at a center-of-mass energy of \(\sqrt{s} = 14 ~\mathrm{TeV}\).

\begin{table}[H]
\caption{The required luminosity to achieve 5$\sigma$ significance $\mathcal{L}_{5\sigma}^{\text{Req}}$ over the SM background for three $M_{A^{\prime}}$ values, taking coupling constants \(\texttt{g}_{\eta} = 0.125\) and \(\texttt{g}_{D} = 1.0\), and $M_{\chi} = 500$ GeV, at $\sqrt{s} = 14 ~\mathrm{TeV}$.}
\centering
\begin{tabular}{|c|c|c|}
\hline
$M_{TS}$ (GeV) & 4000 & 5000 \\
\hline
 $M_{A^{\prime}}$ (GeV)& $\mathcal{L}_{5\sigma}^{\text{Req}}\ (\mathrm{fb}^{-1})$ & $\mathcal{L}_{5\sigma}^{\text{Req}}\ (\mathrm{fb}^{-1})$ \\
\hline
200 & 500 & -\\
300 & 177 & 940  \\
400 & 160 & 600\\
\hline
\end{tabular}
\label{tab:5_sigma}
\end{table}
\subsection{Statistical interpretation}
The shape-based analysis makes effective use of the distributions of $\text{cos}(\theta_{CS})$ as key discriminators. These distributions are especially valuable because they highlight a characteristic signal pattern associated with a typical spin-2 boson, which starkly contrasts the backgrounds outlined by the Standard Model.
An ad-hoc flat 10\% uncertainty is applied to cover all possible systematic effects.

We conducted a statistical test using the profile likelihood method to interpret our results statistically. We utilized the modified frequentist construction CLs \cite{R58, R59} in the asymptotic approximation \cite{R2} to establish exclusion limits on the product of signal cross sections and the branching fraction Br($A^{\prime}$ $\rightarrow \mu\mu$) at a 95\% confidence level. 

\begin{figure*}
\centering
\subfigure[]{
  \includegraphics[width=75mm]{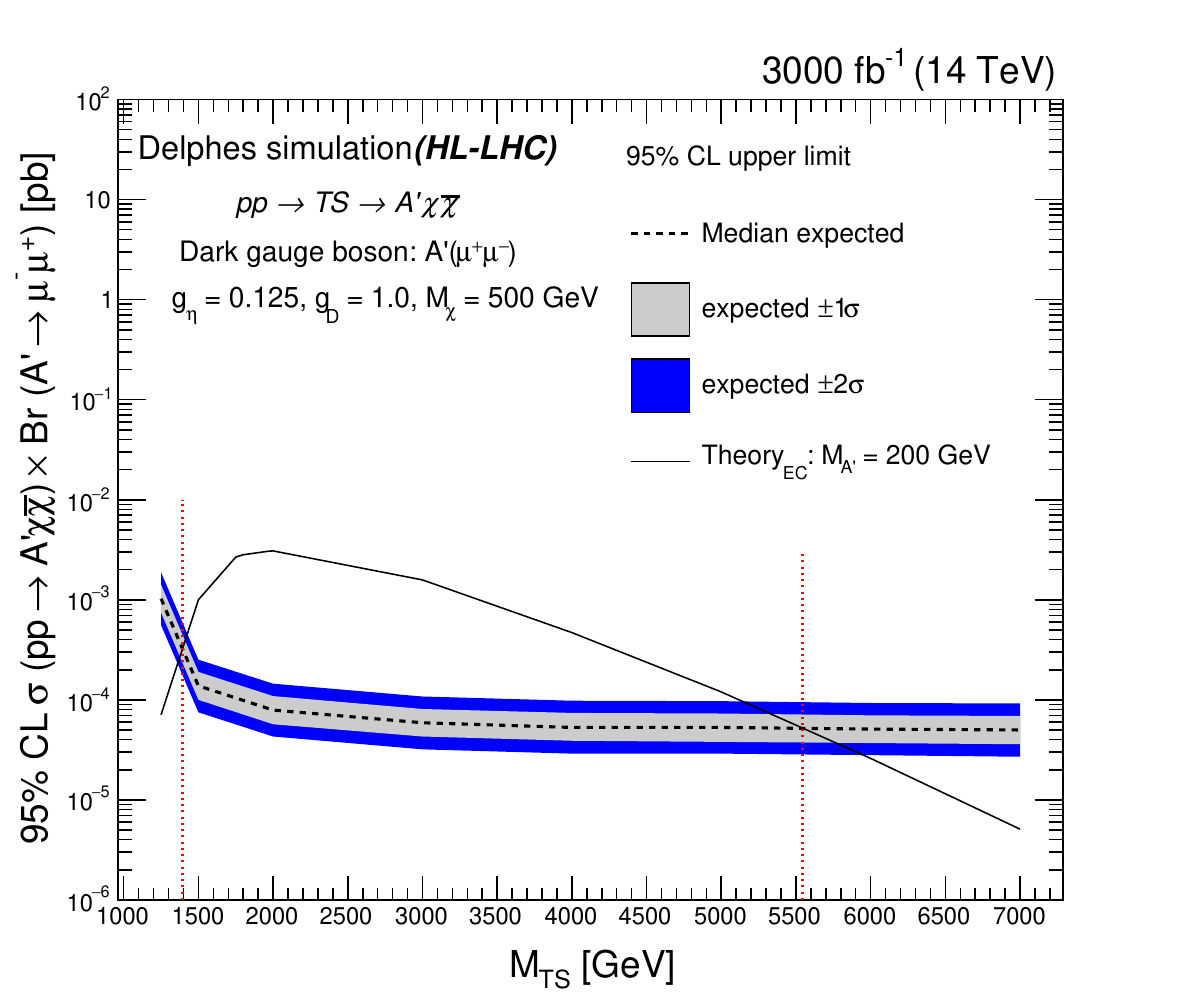}
  \label{limit200}
}
\subfigure[]{
  \includegraphics[width=75mm]{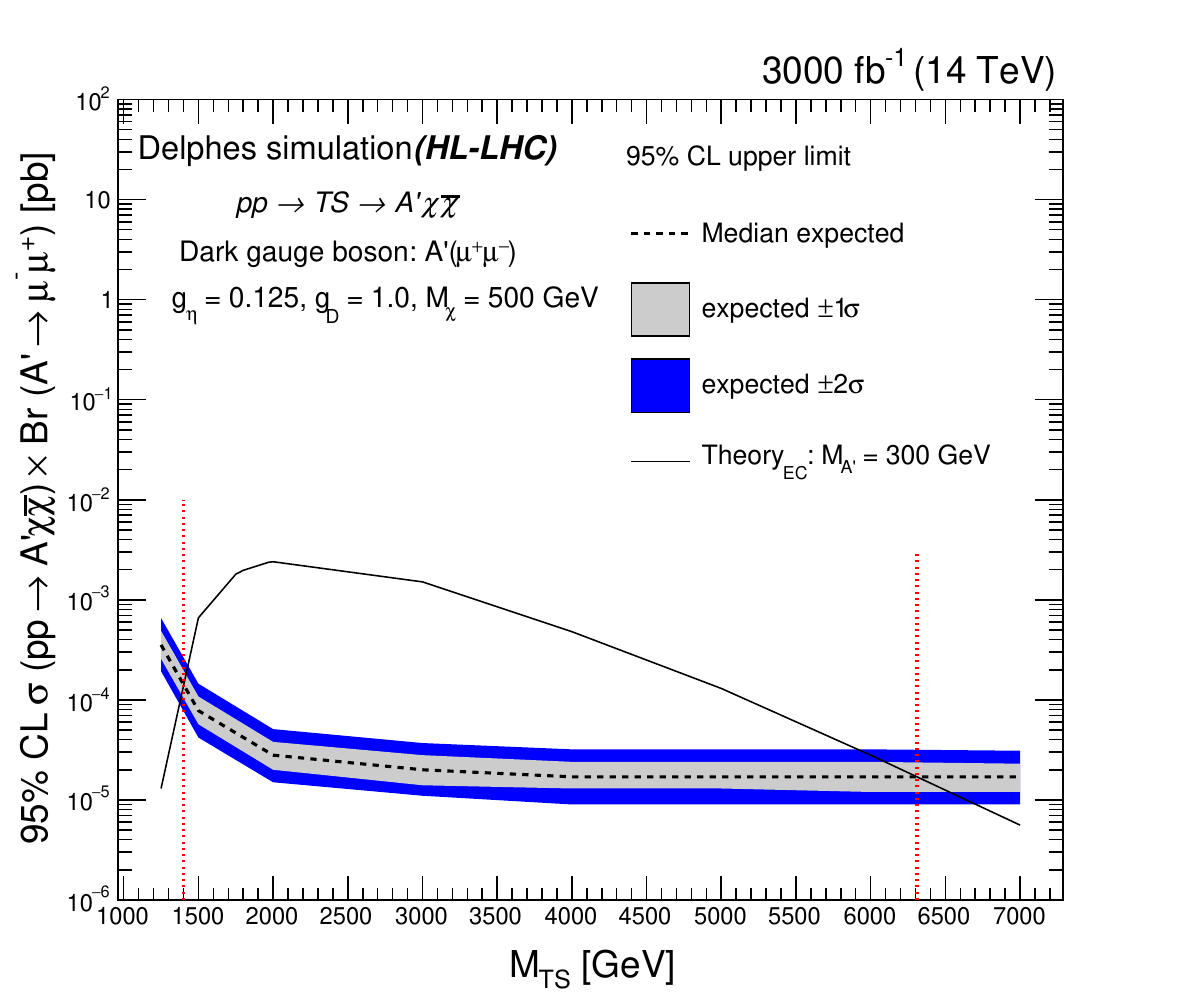}
  \label{limit300}
}
\subfigure[]{
  \includegraphics[width=75mm]{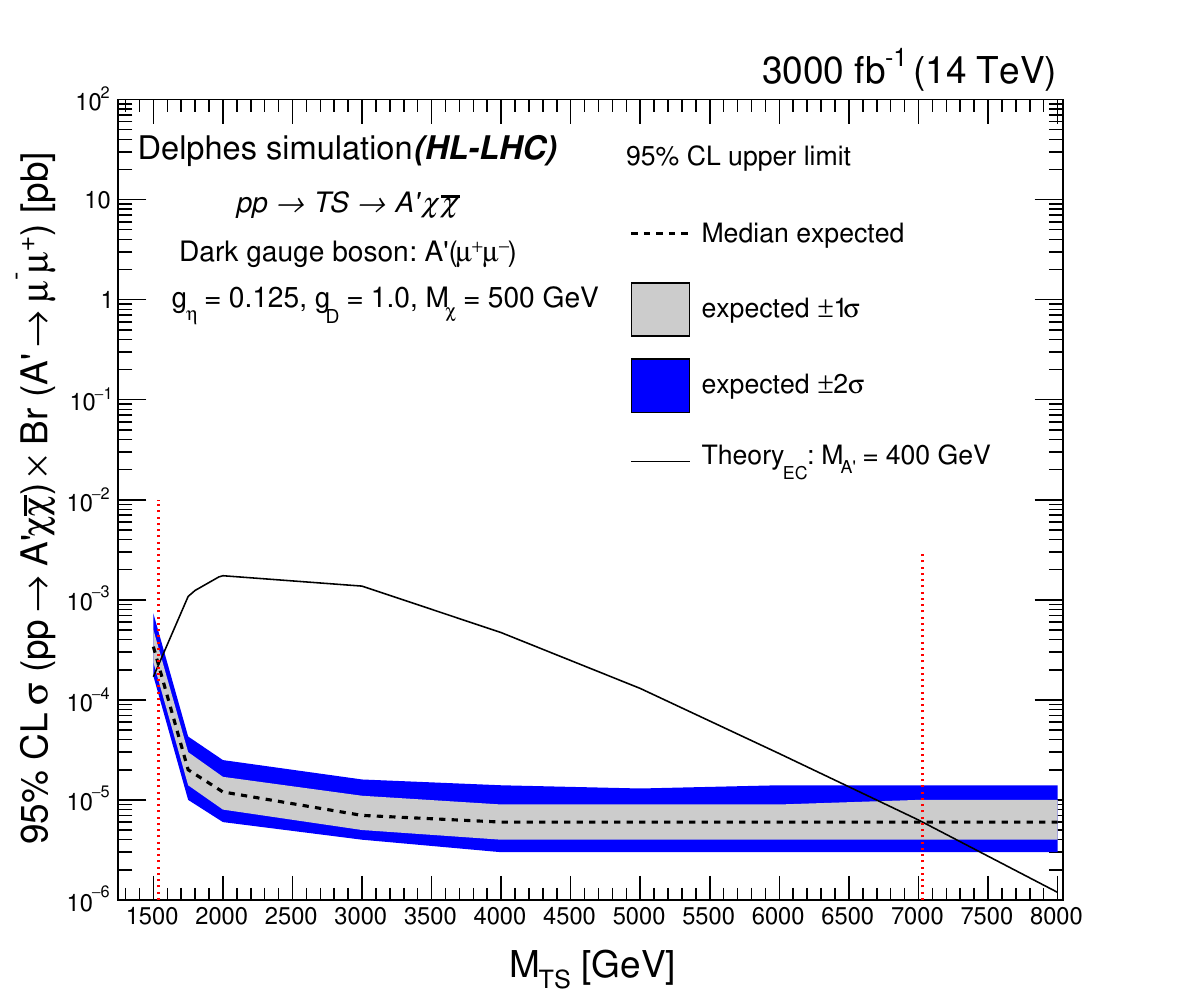}
  \label{limit400}
}
\subfigure[]{
  \includegraphics[width=75mm]{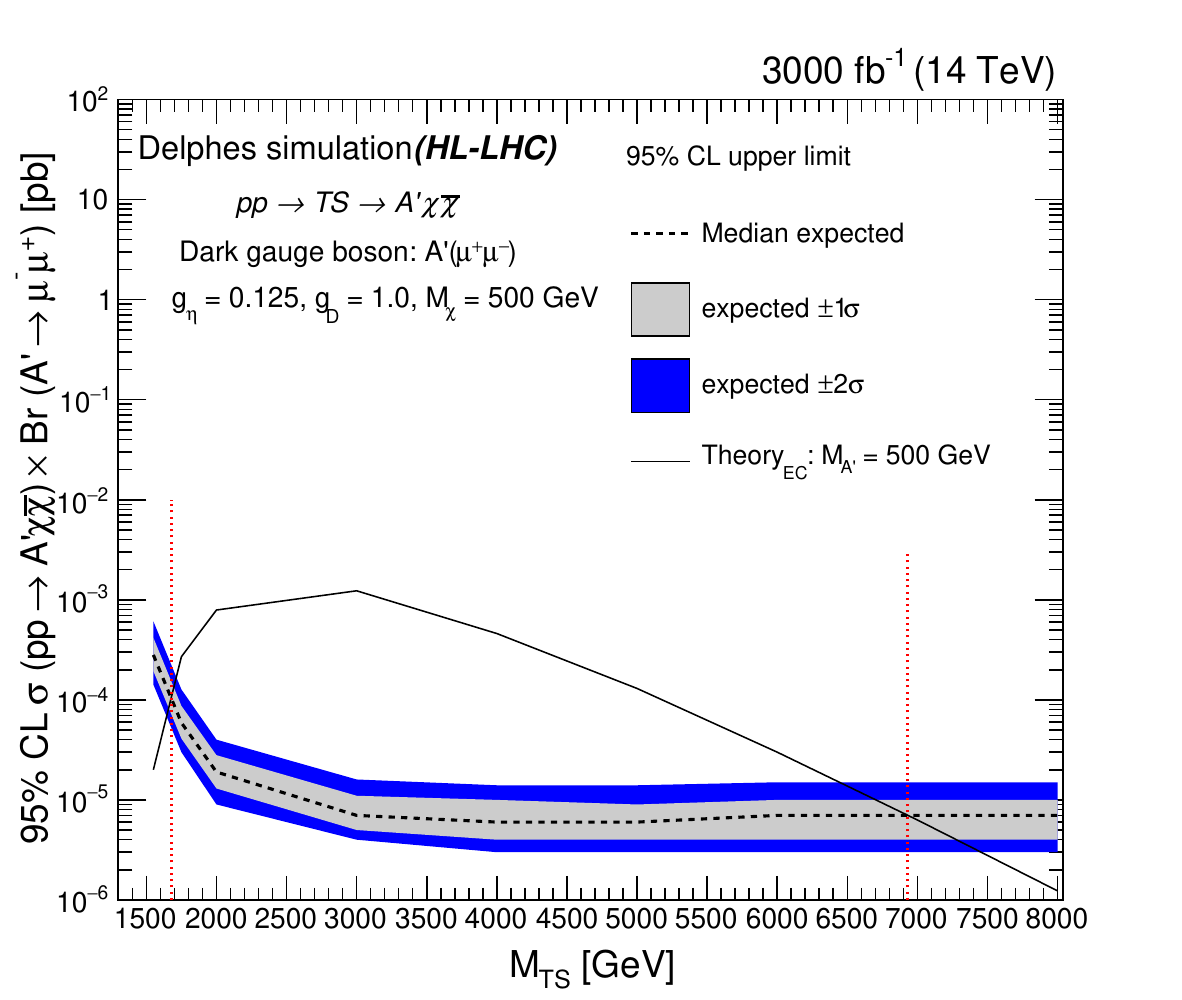}
  \label{limit500}
}
\caption{95\% CL upper limits on the cross-section times the branching ratio (expected), as a function of the mediator’s mass ($M_{TS})$, in the framework of the Einstein-Cartan (EC) gravity, with the muonic decay of the A$^{\prime}$ with $M_{A'}$ = 200 GeV in \ref{limit200}, 300 GeV in \ref{limit300}, 400 GeV in \ref{limit400}, and 500 GeV in \ref{limit500}. 
The black solid curves represent the model based on Einstein-Cartan theory at fixed dark matter mass ($M_{\chi} = 500$ GeV), $\texttt{g}_{\eta} = 0.125$, $\texttt{g}_{D} = 1.0$, and different values of A$^{\prime}$ mass. The vertical red dotted lines in each graph indicate the limit values.}
  \label{limit}
\end{figure*}

In Figure \ref{limit}, we present the expected 95\% confidence level (CL) upper limits on the product of the cross-section times the branching ratio as a function of the mediator’s mass ($M_{TS}$). This analysis is based on Einstein-Cartan theory and focuses on the muonic decay of the A$^{\prime}$. 
The solid black curves represent this model, which features a fixed dark matter mass of \(M_{\chi} = 500\) GeV, coupling parameters \(\texttt{g}_{\eta} = 0.125\) and \(\texttt{g}_{D} = 1.0\) are applied, illustrating a range of A$^{\prime}$ mass values. Specifically, the values for \(M_{A^{\prime}}\) are set at 200 GeV \ref{limit200}, 300 GeV \ref{limit300}, 400 GeV \ref{limit400}, and 500 GeV \ref{limit500}.
The vertical red dotted lines in each graph, shown in Figure \ref{limit}, represent the limit values.

When the mass $M_{A^{\prime}}$ exceeds 500 GeV, there are not enough standard model background events to conduct a meaningful statistical analysis.

\section{Summary}
\label{section:Summary}
One effective strategy for discovering new physics beyond the Standard Model at the Large Hadron Collider involves examining changes in the dilepton mass spectrum, particularly at high masses. Furthermore, analyzing the angular distributions of leptons could distinguish between different models of new physics and assess the spin, parity, and couplings of any potential signals. 

Our study focused on the angular distributions of high-mass dimuon pairs in the Collins-Soper frame. The simulated data we worked with were derived from private proton-proton simulated collisions anticipated at the High-Luminosity LHC (HL-LHC), featuring a center-of-mass energy of \( \sqrt{s} = 14 \) TeV and corresponding to an integrated luminosity of 3000 fb\(^{-1}\). 

We concentrated on the cos\( \theta_{CS} \) variable to differentiate between the Standard Model (SM) background and potential new physics beyond it. Our analysis involved comparing the expected distributions of the cos\( \theta_{CS} \) variable for high-mass SM background events with those from a simplified new physics model based on Einstein-Cartan theory. The resulting signal shape reveals a distinctive feature of a spin-2 A$^{\prime}$ neutral gauge boson, characterized by a symmetric distribution of cos\( \theta_{CS} \) around zero.

To improve the distinction between signal events and standard model (SM) backgrounds, we implemented strict discrimination cuts that effectively diminished the impacts of Drell-Yan and ZZ backgrounds. This approach also significantly curtailed contributions from \( t\bar{t} \), tW, WW, and WZ backgrounds. 

With the application of these strong cuts, a mass \( M_{A^{\prime}} \) of 200 GeV allows us to achieve a 5$\sigma$ discovery at an integrated luminosity of 500 fb\(^{-1}\) for the torsion mass (\( M_{TS} = 4000 \) GeV). In contrast, discovering a torsion mass of \( M_{TS} = 5000 \) GeV necessitates an integrated luminosity exceeding 2000 fb\(^{-1}\). For a mass \( M_{A^{\prime}} \) of 400 GeV, we can reach a 5$\sigma$ discovery with an integrated luminosity of 160 fb\(^{-1}\) when \( M_{TS} = 4000 \) GeV. However, if we consider \( M_{TS} = 5000 \) GeV, a 5$\sigma$ discovery is achievable with an integrated luminosity of 600 fb\(^{-1}\).

Finally, we have established a 95\% confidence level upper limit on certain model-independent parameters within the framework of Einstein-Cartan theory. With coupling values set at \( \texttt{g}_{\eta} = 0.125 \) and \( \texttt{g}_{D} = 1.0 \), along with a dark matter mass of \( M_{\chi} = 500 \) GeV, we determined specific mass exclusions for the torsion field \( M_{TS} \):

- For an A$^{\prime}$ boson mass of 200 GeV, the mass values of \( M_{TS} \) fall between 1396 and 5545.

- For an A$^{\prime}$ boson mass of 300 GeV, the corresponding mass values of \( M_{TS} \) range from 1402 to 6310.

- For an A$^{\prime}$ boson mass of 400 GeV, the mass values of \( M_{TS} \) span from 1537 to 7026.

- For an A$^{\prime}$ boson mass of 500 GeV, the mass values of \( M_{TS} \) extend from 1677 to 6927.

\begin{acknowledgments}
The author of this paper would like to thank Cao H. Nam, the author of \cite{R1}, for his useful discussions about the theoretical models, and for sharing with us the Universal FeynRules Output (UFO) for the model that was used for the generation of the events. 
This paper is based on works supported by the Science, Technology, and Innovation Funding Authority (STDF) under grant number 48289.
\end{acknowledgments}

\textbf{Data Availability Statement:} This manuscript has no associated data or the data will not be deposited. 

\nocite{*}

\end{document}